\documentclass[12pt,ifthen,a4paper]{article}
\usepackage{graphics,rotating,color,pstricks,epsfig,lscape}
\usepackage{amssymb,amsmath,graphicx}

\def\MeV{\ifmmode {\mathrm{\ Me\kern -0.1em V}}\else
                   \textrm{Me\kern -0.1em V}\fi}
\def\GeV{\ifmmode {\mathrm{\ Ge\kern -0.1em V}}\else
                   \textrm{Ge\kern -0.1em V}\fi}
\def\EE{\mathrm{e^+e^-}}
\def\ee{\mathrm{e^+e^-}}
\def\eeHA{\mathrm{e^+e^-\to HA}}
\def\eehA{\mathrm{e^+e^-\to hA}}

\def\HAbbbb{\mathrm{HA\to b\bar{b}b\bar{b}}}
\def\HAbbtt{\mathrm{HA\to b\bar{b}\tau^+\tau^-}}
\def\HAttbb{\mathrm{HA\to\tau^+\tau^- b\bar{b}}}
\def\HAbbttbb{\mathrm{HA\to b\bar{b}\tau^+\tau^-,\tau^+\tau^- b\bar{b}}}

\def\sHAbbbb{\mathrm{\sigma(\ee\rightarrow HA\rightarrow b\bar{b}b\bar{b})}}
\def\sHAbbtt{\mathrm{\sigma(\ee\rightarrow HA\rightarrow b\bar{b}\tau^+\tau^-)}}
\def\sHAttbb{\mathrm{\sigma(\ee\rightarrow HA\rightarrow \tau^+\tau^- b\bar{b})}}
\def\sHA{\mathrm{\sigma_{HA}}}
\def\sHZSM{\mathrm{\sigma_{HZ}^{SM}}}
\def\bbbb{\mathrm{b\bar{b}b\bar{b}}}
\def\bbtt{\mathrm{b\bar{b}\tau^+\tau^-}}
\def\ghAZ{\mathrm{g_{hAZ}}}
\def\gHAZ{\mathrm{g_{HAZ}}}
\def\sqrts{\sqrt{s}}
\def\WW{\mathrm{\EE\to WW}}
\def\ZZ{\mathrm{\EE\to ZZ}}
\def\Wen{\mathrm{\EE\to We\nu}}
\def\Zee{\mathrm{\EE\to Z\EE}}
\def\tt{\mathrm{\EE\to t\bar{t}}}
\def\qq{\mathrm{\EE\to q\bar{q}}}
\def\qqal{\mathrm{q\bar{q}(q=u,d,s,c,b)}}
\def\WWqqqq{\mathrm{WW\to q\bar{q}q\bar{q}}}
\def\WWqqln{\mathrm{WW\to q\bar{q}\ell\nu_{\ell}}}
\def\ttWbWb{\mathrm{t\bar{t}\to W^+bW^-\bar{b}}}
\def\ZZqqqq{\mathrm{ZZ\to q\bar{q}q\bar{q}}}
\def\ZZqqnn{\mathrm{ZZ\to q\bar{q}\nu\bar{\nu}}}
\def\ZZqqll{\mathrm{ZZ\to q\bar{q}\ell^+\ell^-}}
\def\Zeeqq{\mathrm{Z\EE\to q\bar{q}\EE}}
\def\Wenqq{\mathrm{We\nu\to q\bar{q} e\nu}}
\def\tanb{\tan\beta}
\def\mH{m_\mathrm{H}}
\def\mA{m_\mathrm{A}}
\def\Evis{\mathrm{E_{vis}}}
\def\qq{\mathrm{q\bar{q}}}

\def\eeqq{\mathrm{\EE\to \qq}}
\def\Nefo{\mathrm{N_{enflow}}}
\def\Bh{\mathrm{B_1}}
\def\Bl{\mathrm{B_2}}

\def\mumu{\mathrm{\mu^-\mu^+}}
\def\LHA{\mathrm{L_{HA}}}

\def\BrHbb{\mathrm{Br(H\to b\bar{b})}}
\def\BrAbb{\mathrm{Br(A\to b\bar{b})}}
\def\BrHtt{\mathrm{Br(H\to \tau\tau)}}
\def\BrAtt{\mathrm{Br(A\to \tau\tau)}}
\def\gH{\mathrm{\Gamma_H}}
\def\gA{\mathrm{\Gamma_A}}

\def\ra{\rightarrow}

\begin{document}
\begin{titlepage}
\begin{flushright}
{\bf LC-PHSM-2004-006}  \\ {\bf March 2004  }
\end{flushright}
\vspace{1 cm}
\begin{center}{\bf\Large  Study of Higgs Boson Pair Production
       at Linear Collider}
\end{center}
\vspace{0.8 cm}
\begin{center}
  {\large\bf  K.~Desch$^{a}$, T.~Klimkovich $^{a,b}$, T.~Kuhl$^{b}$,}
  ~{\large\bf   A.~Raspereza$^{b}$~  }\\
\vspace{0.2 cm}
{\em $^a$ Universit\"at Hamburg, Institut f\"ur Experimentalphysik \\
Notkestrasse 85, D-22607 Hamburg, Germany.}\\
{\em $^b$ DESY, Notkestrasse 85, D-22607 Hamburg, Germany.}\\
\end{center}
\vspace{4mm}

\begin{abstract}
We study the potential of the TESLA linear collider operated at 
a center-of-mass energy of 500 to 1000$\GeV$ 
for the measurement of the neutral Higgs boson 
properties within the framework of the MSSM. The process of associated 
Higgs boson production with subsequent decays of Higgs 
bosons into b-quark and $\tau$-lepton pairs is considered.
An integrated luminosity of 500 fb$^{-1}$ is assumed at each energy. 
The Higgs boson masses and production cross sections 
are measured by reconstructing the $\bbbb$ and $\bbtt$ final states. 
The precision of these measurements 
is evaluated in dependence of the Higgs boson masses. 
Under the assumed experimental conditions
a statistical accuracy ranging from 0.1 to 1.0 GeV is achievable
on the Higgs boson mass.
The topological cross section $\sHAbbbb$ can be determined with  
the relative precision of 1.5 -- 6.6 \% and cross sections $\sHAbbtt$ and $\sHAttbb$ with precision of 4 -- 30 \%.
Constraints on the Higgs boson widths can be set exploiting
$\bbtt$ channel. The 5$\sigma$ discovery limit corresponds to the Higgs mass of around 385 GeV for the degenerate Higgs boson masses in 
the $\HAbbbb$ channel at $\sqrts$ = 800 GeV with integrated luminosity
of 500 fb$^{-1}$. The potential of the Higgs mass 
determination for the benchmark point SPS 1a for the process $\HAbbbb$ 
at $\sqrt{s}$ = 1 TeV and luminosity 1000 fb$^{-1}$ is investigated.

\end{abstract}
\end{titlepage}
%
%
\section{Introduction}

Elucidating the mechanism of electroweak symmetry breaking will be one of the central 
tasks at a future linear $\EE$ collider. In the minimal 
Standard Model (SM) breaking of electroweak symmetry 
is provided by one Higgs doublet, which introduces an 
additional spin-0 particle, the Higgs boson~\cite{higgs}. 
Among models with an extended Higgs sector, two Higgs doublet 
models (2HDM)~\cite{higgs_hunters} are of particular
interest, since this structure of the Higgs sector is required 
in the Minimal Supersymmetric Standard Model (MSSM)~\cite{mssm}, the 
most seriously considered extension of the SM. 
The Higgs sector of the CP-conserving MSSM comprises the
following physical states:
two CP-even Higgs bosons, the lighter of which is denoted h and
the heavier H, one CP-odd Higgs boson A and two charged 
bosons H$^\pm$. Two crucial parameters, 
which influence the Higgs boson phenomenology, are $\tanb$, the ratio of vacuum expectation values
of the two Higgs doublets, and the angle $\alpha$, which 
controls mixing in the CP-even sector. 

The cross sections of the Higgs-strahlung 
and fusion processes, involving CP-even Higgs bosons,
\begin{equation}
\mathrm{
e^+e^- \to hZ(HZ), \hspace{7mm}
e^+e^- \to h(H)\nu_e\bar{\nu}_e, \hspace{7mm}
e^+e^- \to h(H)e^+e^-,
}
\label{eq:fusion}
\end{equation}
are defined in terms of the strength of the Higgs couplings to 
weak bosons at tree level. These depend on the parameters $\alpha$ 
and $\beta$ in the following way:
\begin{equation}
\mathrm{
g_{hZZ,hWW} \sim \sin(\beta-\alpha), \hspace{2cm}
g_{HZZ,HWW} \sim \cos(\beta-\alpha).
}
\end{equation}
In the MSSM, the set of the tree level couplings between
neutral Higgs particles and weak bosons is extended by 
two additional couplings,
\begin{equation}
\mathrm{
g_{hAZ} \sim \cos(\beta-\alpha),\hspace{2cm}
g_{HAZ} \sim \sin(\beta-\alpha).
}
\end{equation} 
As a consequence, in $\EE$ collisions, the Higgs-strahlung 
and fusion processes will be complemented by the associated 
Higgs boson pair production mechanisms:
\begin{equation}
\mathrm{
e^+e^- \to hA(HA).
}
\label{eq:hA}
\end{equation}

To fully explore the MSSM Higgs sector, the processes (\ref{eq:hA}) 
must be studied, since 
they directly probe $\ghAZ$ and $\gHAZ$ trilinear couplings,
provide access to the A boson properties
and supplement the study of the CP-even Higgs boson
properties via Higgs-strahlung and fusion mechanisms.

The study of the Higgs pair production process is motivated
by a so-called decoupling limit of the MSSM in which 
the h boson approaches the properties of the SM Higgs 
boson. The closer MSSM scenario moves towards the decoupling
limit the more difficult it becomes to distinguish the Higgs sector
from the SM. In such a scenario the detection of heavier neutral 
Higgs bosons would be crucial for establishing an extended Higgs sector.
The decoupling limit is approached relatively fast for  
large values of the H and A boson masses, $\mA,\mH$ 
$>$ 200 GeV, in a large portion of the MSSM parameter space.
The distinct feature of this scenario is vanishing coupling of heavy CP-even
Higgs boson to weak bosons, $\cos(\beta-\alpha)\ra 0$.
As a consequence, the H boson production via
the fusion and Higgs-strahlung processes is significantly 
suppressed, whereas the cross section of the $\eeHA$ process
reaches its maximal value making associated heavy Higgs pair
production a promising channel for the detection of the H and A 
bosons at a future linear $\ee$ collider. 
It should be also emphasized that in the decoupling limit the H and A 
bosons are almost degenerate in mass and have similar decay 
properties\footnote{
These features will be exploited later on in our analysis.}. 

In this note the study of Higgs boson pair production   
at a future linear $\ee$ collider is presented. This analysis 
extends the previous studies~\cite{tdr,tdr_bbbb}. The analyzed topologies 
include the $\bbbb$ and $\bbtt$ final states.
In the following we will assume the $\eeHA$ production process.
However, this assumption does not restrict the generality 
of our study and the analyses developed in this 
note are applicable to the process $\eehA$ as well as to the processes 
of associated Higgs pair production in CP-violating MSSM scenarios 
where the neutral Higgs bosons are no longer CP eigenstates~\cite{Pilaftsis}.

This note is organized as follows.
In Section 2 the assumed experimental conditions and 
main features of the detector are discussed. 
Section 3 describes the simulation of the signal and the 
relevant background processes.
In Section 4 the analysis tools, employed in our study, are briefly 
reviewed. Section 5 describes the analysis
of the $\bbbb$ and $\bbtt$ final states.
In Section 6, we discuss the prospects of 
measuring the Higgs boson properties using
the $\bbbb$ and $\bbtt$ topologies. In Appendix the interpretation of this analysis is done for the SPS 1a benchmark point. 

%
%
\section{Experimental Conditions and Detector \newline Simulations}

The study is performed for a linear collider operated 
at a center-of-mass energies of 500$\GeV$ to 1000$\GeV$ and  
an event sample corresponding to an integrated luminosity
of 500 fb$^{-1}$. 

The detector used in the simulation follows the proposal
for the TESLA collider presented in 
the TDR~\cite{tdr}. Studies were also performed in Asia and North America~\cite{Kiyoura:2003tg,orange_book}.
The interaction region is surrounded by a tracker consisting
of a multi-layered pixel micro-vertex detector (VTX) as the innermost part 
and a large time projection chamber (TPC) supplemented with additional silicon strip detector in the forward region (FTD) and two bands between VTX and TPC. In radial direction follow
an electromagnetic calorimeter, a hadron calorimeter,
the coils of a superconducting magnet
and an instrumented iron flux return yoke.
The solenoidal magnetic field is 4 Tesla.

The envisaged impact parameter resolution 
provided by the vertex detector is 

\begin{equation}
  \delta(IP_{r\phi,z})\le 5\mu m \oplus \frac{10\mu m\hspace{1mm}GeV/c}{p\sin^{3/2}\theta},  
\end{equation}
where $p$ is the track momentum and $\theta$ is the polar angle.
The overall central tracker momentum resolution is 
\begin{equation}
  \sigma \left(\frac{1}{p_t}\right) \le 5 \cdot 10^{-5} ~ [\GeV/c]^{-1},
\end{equation}
where $p_t$ is the transverse momentum. To enable efficient separation
of neutral and charged particles, highly granular 
electromagnetic and hadronic calorimeters are foreseen for the 
TESLA detector. Their energy resolutions are:
\begin{equation}
\frac{\sigma_{E_{el}}}{E_{el}} = \frac{10\%}{\sqrt E_{el}} \oplus 1 \%,\hspace{2cm}
\frac{\sigma_{E_{h}}}{E_{h}} = \frac{50\%}{\sqrt E_{h}} \oplus 4 \%,
\end{equation}
where $E_{el}$ and $E_{h}$ are the energies
measured for electrons and hadrons in the corresponding
calorimeters. The polar angle coverage of the central tracker 
maintaining the resolution is
$|\cos \theta| < 0.85$, above this range the tracking 
resolution deteriorates. The electromagnetic and hadron 
calorimeters cover $|\cos \theta| < 0.996$ 
maintaining the resolution over the whole angular range.

The detector response is simulated with parametric fast 
simulation program SIMDET~\cite{simdet}. 
Beamstrahlung is accounted for using CIRCE~\cite{circe}.  

The event reconstruction is done in terms of energy flow objects. 
First, tracks are measured with tracking system and associated 
to calorimetric clusters to define charged energy flow objects 
(electrons, muons and charged hadrons). Since momentum measurement
with tracking system is much more accurate than direction and energy
measurements with calorimeters, the tracking information
is used for estimation of the four-momentum of the charged objects.  
Calorimetric clusters with no associated tracks are regarded as 
neutral energy flow objects (photons and neutral hadrons). 
Measurements of the four-momentum of neutral objects are solely
based on the calorimetric information.

%
%
\section{Physics Processes and Monte Carlo \newline  Samples}

Samples of $\eeHA$ events are generated 
for several Higgs boson mass hypotheses 
with PYTHIA 6.2~\cite{pythia}, including initial state radiation.
The tree level cross section of the HA production 
is related to the Higgs-strahlung cross section in the SM,
$\sHZSM$, in the following way~\cite{higgs_hunters}:
\begin{equation}
\mathrm{
 \sHA = g^2_{HAZ}\bar{\lambda}_{HA}\sHZSM,
}
\end{equation}
where 
\begin{equation}
\mathrm{
\bar{\lambda}_{HA}=\frac{\lambda^{3/2}_{HA}}{\lambda^{1/2}_{HZ}[12m^2_Z/s+\lambda_{HZ}]}
}
\end{equation}
accounts for the correct suppression of the P-wave cross section near the kinematic threshold. The quantity $\lambda_{ij}=[1-(m_i+m_j)^2/s][1-(m_i-m_j)^2/s]$ is the usual momentum factor of the two particle phase space. The Higgs-strahlung cross section in the SM is computed using 
the program HPROD~\cite{hprod}. 

For the case of maximal 
allowed $\gHAZ$ coupling, $\sin(\beta-\alpha)$ = 1, which is 
assumed for these studies, signal cross sections are given in Table~\ref{tab:higgs_sample}. 
The branching fractions of the Higgs bosons into b-quarks and $\tau$-lepton 
pairs are set to their typical values in the MSSM : 
$\BrHbb$ = $\BrAbb$ = 90\%, $\BrHtt$ = $\BrAtt$ = 10\%.

For the background estimation, the following processes are generated
using PYTHIA 6.2: $\WW$, $\ZZ$, 
$\eeqq$\footnote{The $\tt$ events are also taken into account},
$\Zee$ and $\Wen$. The cross sections 
for the most important background processes 
are given in Table~\ref{tab:backgrounds}. 


\section{Analysis tools}
\subsection{B-tagging}
\label{label:btag}
Identification of b-quarks plays a crucial role in this analysis. Efficient tagging of jets containing
heavy flavour hadrons will be achieved with a highly granular micro-vertex
detector, allowing for precise reconstruction of track parameters in the 
vicinity of the primary interaction point.

The procedure of tagging b-jets exploits information from the single track as well as secondary vertex information. 
Secondary vertices are searched within jets using the 
package ZVTOP~\cite{ZVTOP} 
developed for the SLD experiment, where tracks are described
as probability tubes and seed vertices are defined as regions 
where these tubes overlap. Afterwards an attempt is made to assign 
additional tracks by an iterative procedure. For each found vertex
the invariant total mass and momentum are calculated from the four-momenta of energy flow
objects assigned to the vertex. Three dimensional decay length and 
decay length significance are also computed.

For a jet flavour separation a neural network is developed. 
The detailed description of the neural network implementation 
for the full simulation of the TESLA detector with the 
BRAHMS program~\cite{Brahms}
can be found in~\cite{XHansen}. Three different neural networks 
are introduced for jets with no, one and more than one secondary 
vertices found. For jets without secondary vertex,
the impact parameter joint probability~\cite{ImpParProb} and the two 
highest impact parameter
significances are used as inputs for the neural network. If jet contains
at least one secondary vertex, an additional information, 
including vertex masses, momenta, decay lengths and decay 
length significances, is fed into the neural network. 

The neural networks are trained on event samples simulated with SIMDET
using the same variables and jet classification as in BRAHMS. For the 
analysis, a jet-wise tag, referred hereafter as jet b-tag variable, 
is used. For a jet with neural network output $x$ it is defined as
\begin{equation}
B(x) = \frac{f_b(x)}{f_b(x)+f_{udsc}(x)},
\end{equation}
where  
$f_b(x)$ and $f_{udsc}(x)$ are probability density functions
of neural network output in a samples of b-- and udsc--jets, respectively.
Tagging of c-jets proceeds in a similar way.

\begin{figure}
\begin{center}
\includegraphics[width=1.1\textwidth]{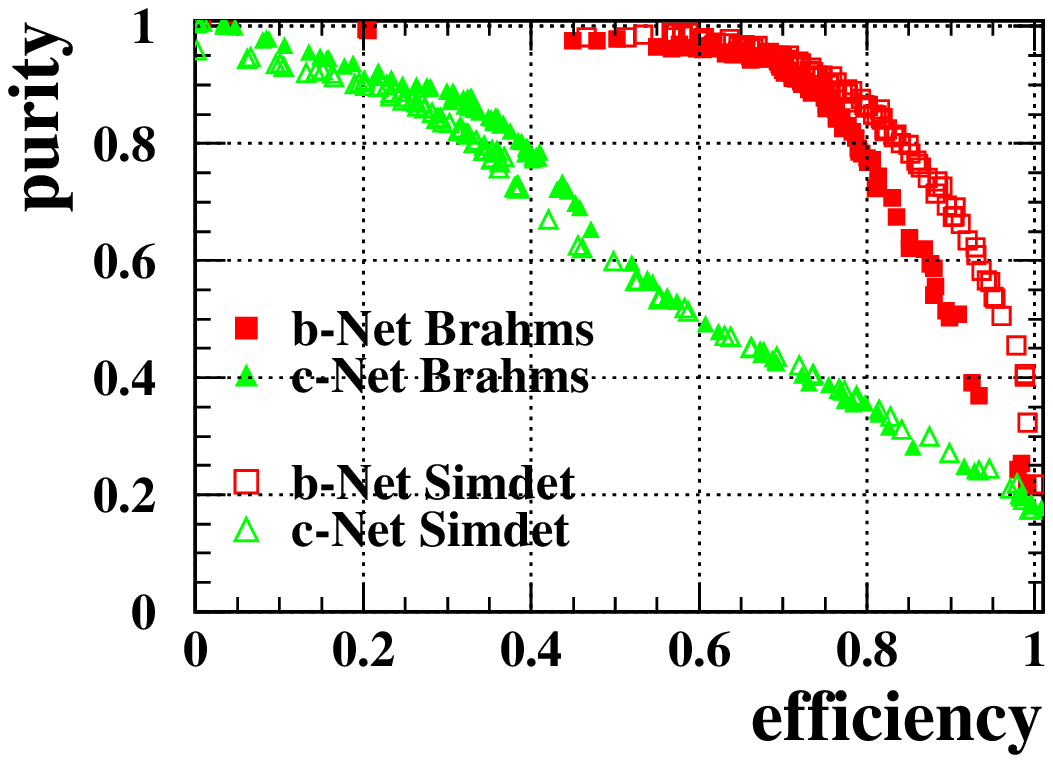}
\caption{b-tag and the c-tag purity of the neural network (Section~\ref{label:btag})  versus efficiency curves for $\mathrm{Z\ra q\bar{q}}$ events simulated with SIMDET and BRAHMS at a center-of-mass energy of $\sqrt{s}=91.2\,\mathrm{GeV}$. Filled squares (triangles) correspond to the neural network output for the b-events (c-events) simulated with BRAHMS. Open squares (triangles) correspond to the neural network output for the b-events (c-events) simulated with SIMDET. The neural networks are trained on event samples simulated with SIMDET
using the same variables and jet classification as in BRAHMS.
\label{fig:Btag}
}
\end{center}
\end{figure}

Fig.~\ref{fig:Btag} shows b-tag and c-tag
purity versus efficiency curves for $\mathrm{Z\ra q\bar{q}}$ 
events simulated at 
a center-of-mass energy of $\sqrt{s}=91.2\,\mathrm{GeV}$.
Results obtained with SIMDET and BRAHMS are compared in this figure.
The c-tag performance agrees within 5\% over the entire range of efficiency. 
Some discrepancy in modelling of the b-tag 
is observed in the region of high efficiency and low purity. 
This discrepancy occurs due to not adequate modelling of the 
resolution tails in the impact parameter joint probability 
distribution by SIMDET. 
However, both $\bbbb$ and $\bbtt$ analyses impose b-tag requirements
strong enough not to be sensitive to this discrepancy.
Some flavour tag related systematic studies have been performed.
It has been found that the b-tag performance is nearly independent of 
the center-of-mass energy in the range from $\sqrt{s}=91.2\,\mathrm{GeV}$ 
to $\sqrt{s}=500\,\mathrm{GeV}$. Furthermore, possible changes in  
micro-vertex detector configuration are found to have 
marginal impact on the b-tag performance. For example,
removing innermost silicon layer changes the selection efficiency by not
more than 5\%. These studies confirm the stability of b-tagging. 
However, c-tag and e.g. b-quark charge tagging are depending on the 
innermost layer.   


\subsection{Kinematic Fit}

The mass resolution of the reconstructed Higgs bosons is improved
by means of a kinematic fit. In the $\bbbb$ analysis, 
conservation of four-momentum is required, leading 
in total to four constraints. In the case of the $\bbtt$ topology, 
the energies of $\tau$-leptons are not measurable due to undetectable
neutrinos. However, the flight direction of highly boosted $\tau$-leptons
can be approximated with good accuracy by the direction of the 
total momentum of its visible decay products. Exploiting 
this approximation together with the the four-momentum conservation 
requirement, we arrive at a two constraint fit. 

\begin{figure}
\begin{center}
\includegraphics[width=0.75\textwidth]{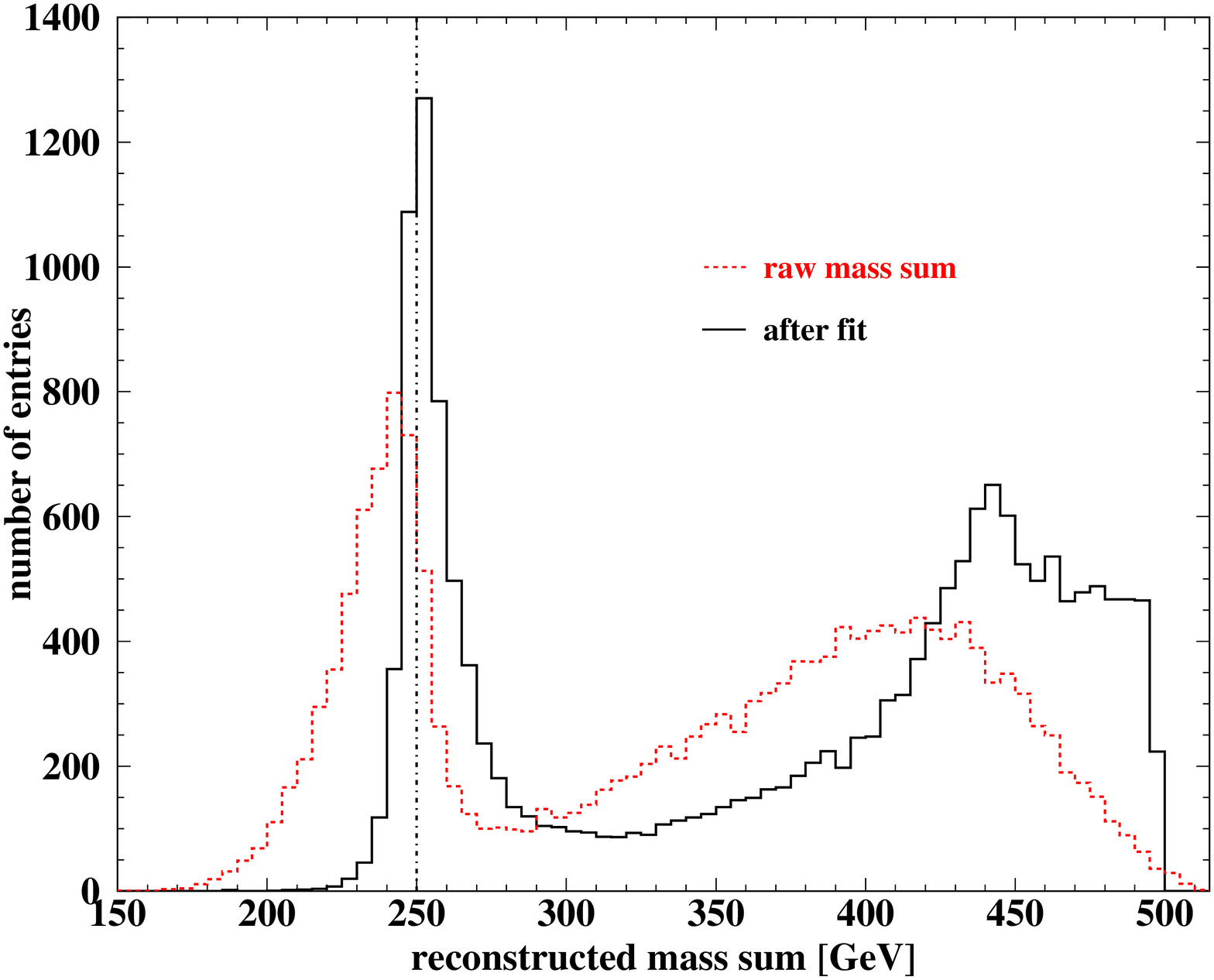} \\
\vspace{-2mm}
\includegraphics[width=0.75\textwidth]{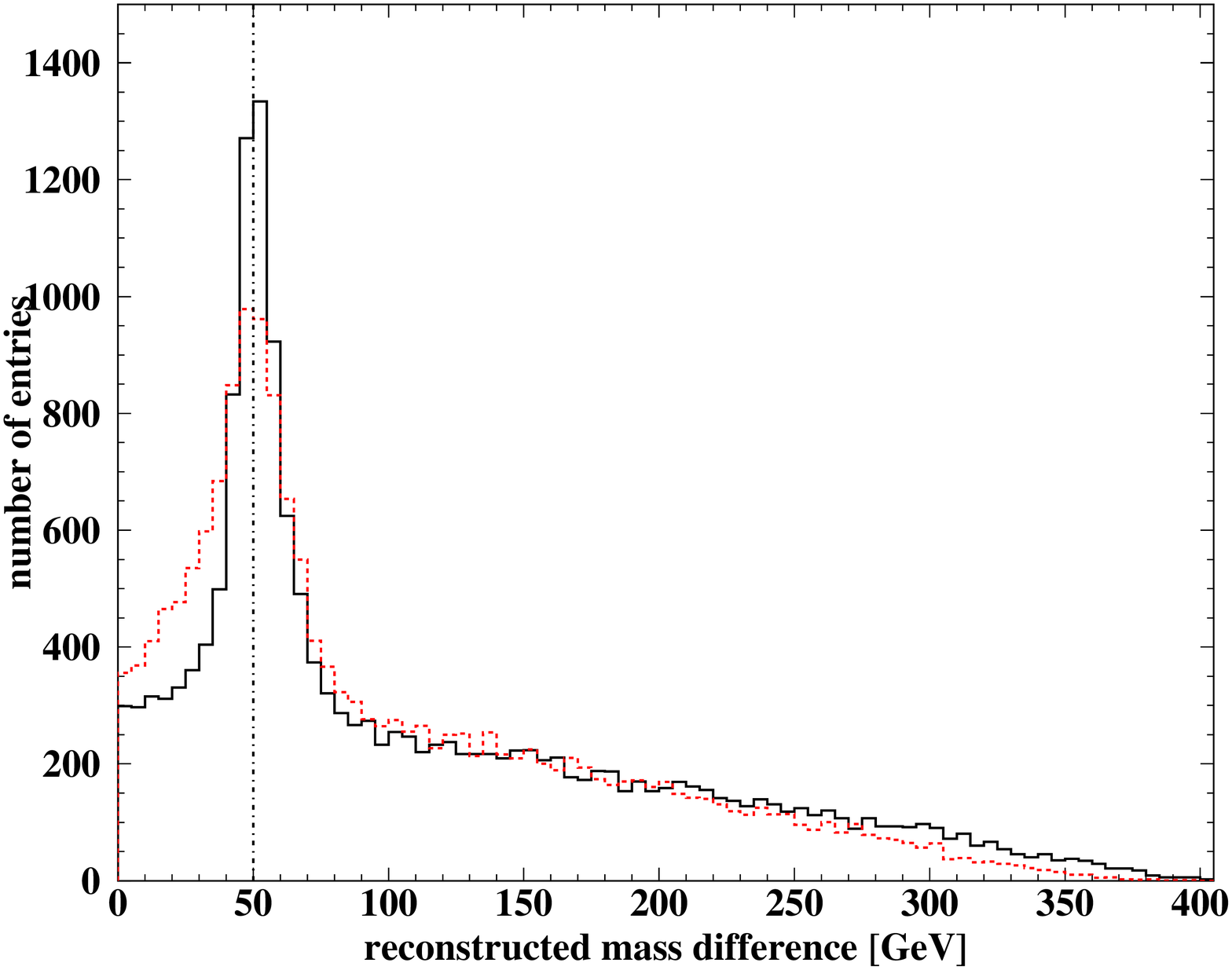}
\caption{Distributions of the reconstructed invariant 
mass sum (upper figu- \newline re) and mass difference (lower figure) 
in the sample of the $\HAbbbb$ events
with ($\mH$,$\mA$) = (150,100) GeV at $\sqrt{s}=500\,\mathrm{GeV}$. Dashed histograms show raw spectra obtained using only measured jet 
angles and energies. Solid histograms show spectra obtained  
after applying 4C kinematic fit. Three entries per event contribute
to distributions, corresponding to three possible di-jet pairings.
\label{fig:kf_bbbb}
}
\end{center}
\end{figure}

\begin{figure}
\begin{center}
\includegraphics*[width=0.75\textwidth]{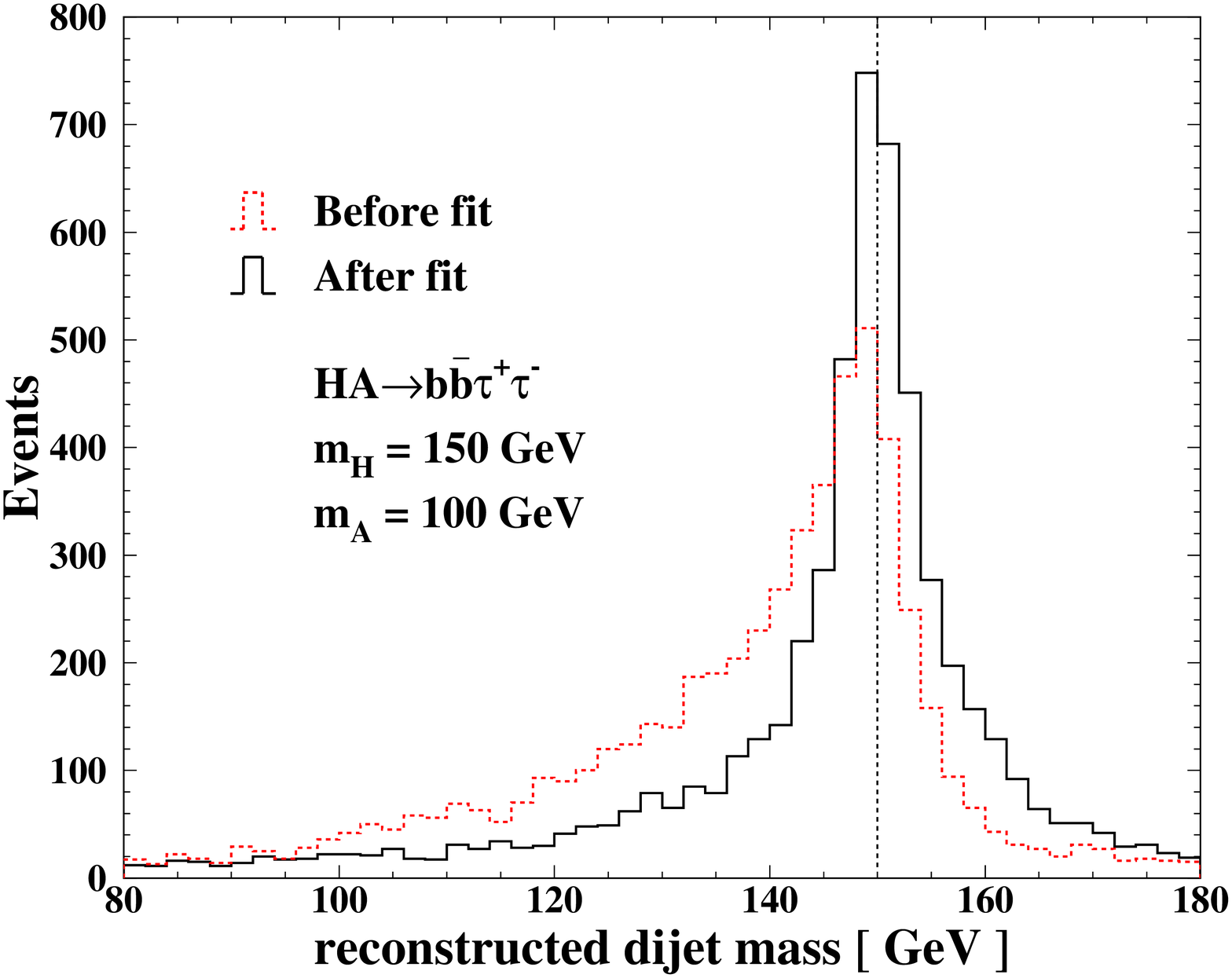} \\
\vspace{-2mm}
\includegraphics*[width=0.75\textwidth]{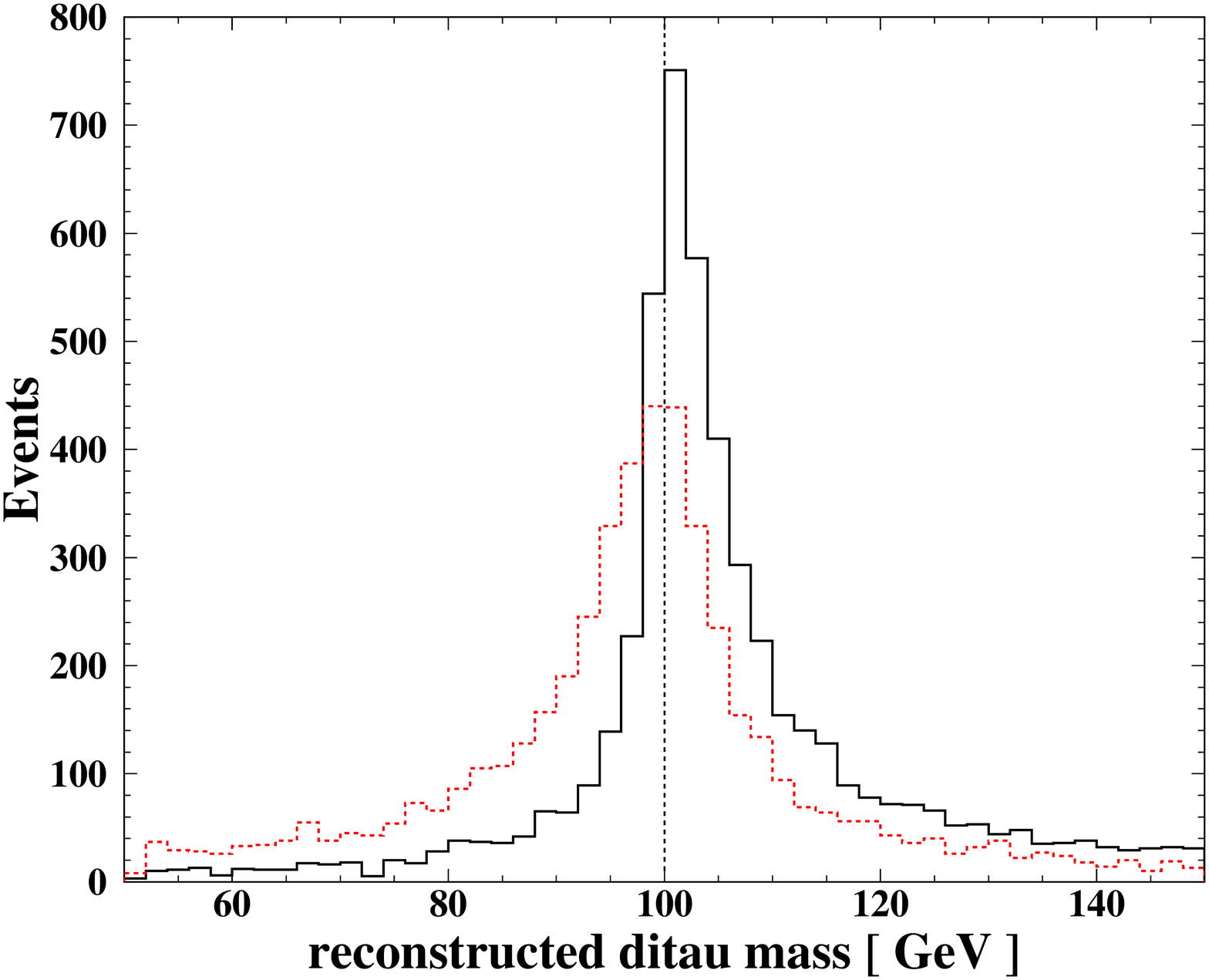} 
\end{center}
\caption{
The mass spectrum of the di-jet (upper figure) and di-tau (lower figure)
systems in the sample of the $\HAbbtt$ events with ($\mH$,$\mA$) = (150,100) GeV at $\sqrt{s}=500\,\mathrm{GeV}$. Dashed histograms 
represent distributions before applying kinematic fit, solid
- after applying kinematic fit.
\label{fig:kf_bbtt}
}
\end{figure}

The jet angular and momentum resolution functions and the 
angular resolution of $\tau$-leptons
are derived from Monte Carlo studies.
The kinematic fit is performed 
using the code developed by DELPHI~\cite{delphi_fit}.
For the samples of the $\HAbbbb$ and $\HAbbtt$ events 
with ($\mH,\mA$) = (150,100)~\GeV, the performance of the kinematic 
fit is illustrated in Fig.~\ref{fig:kf_bbbb} and~\ref{fig:kf_bbtt},
respectively.

\section{Analysis Procedures}

%
%
\subsection{The $\HAbbbb$ Channel}

\begin{figure}[p]
\begin{center}
\includegraphics[width=0.5\textwidth]{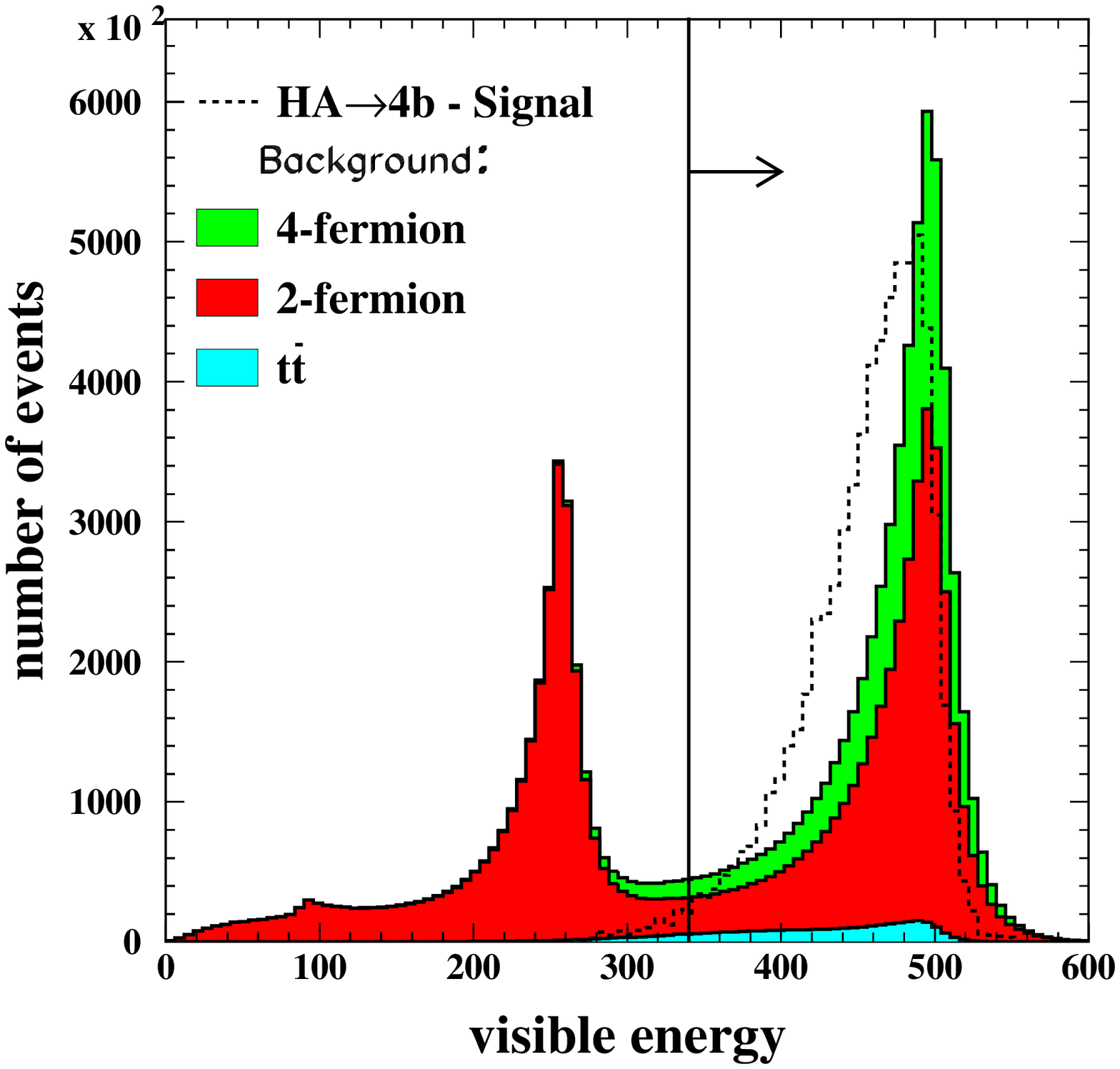}
\hspace{-4mm}
\includegraphics[width=0.5\textwidth]{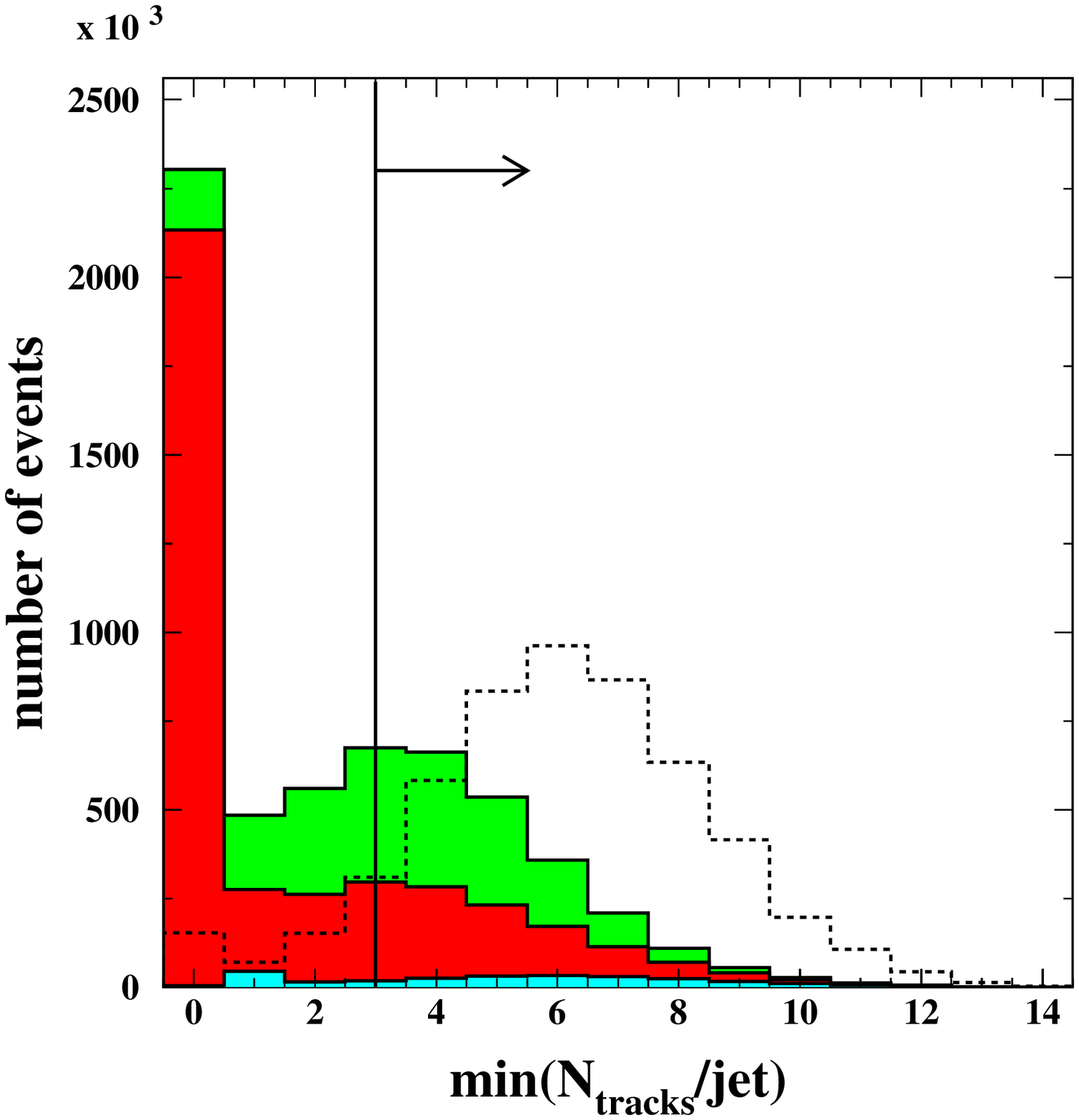}
\includegraphics[width=0.5\textwidth]{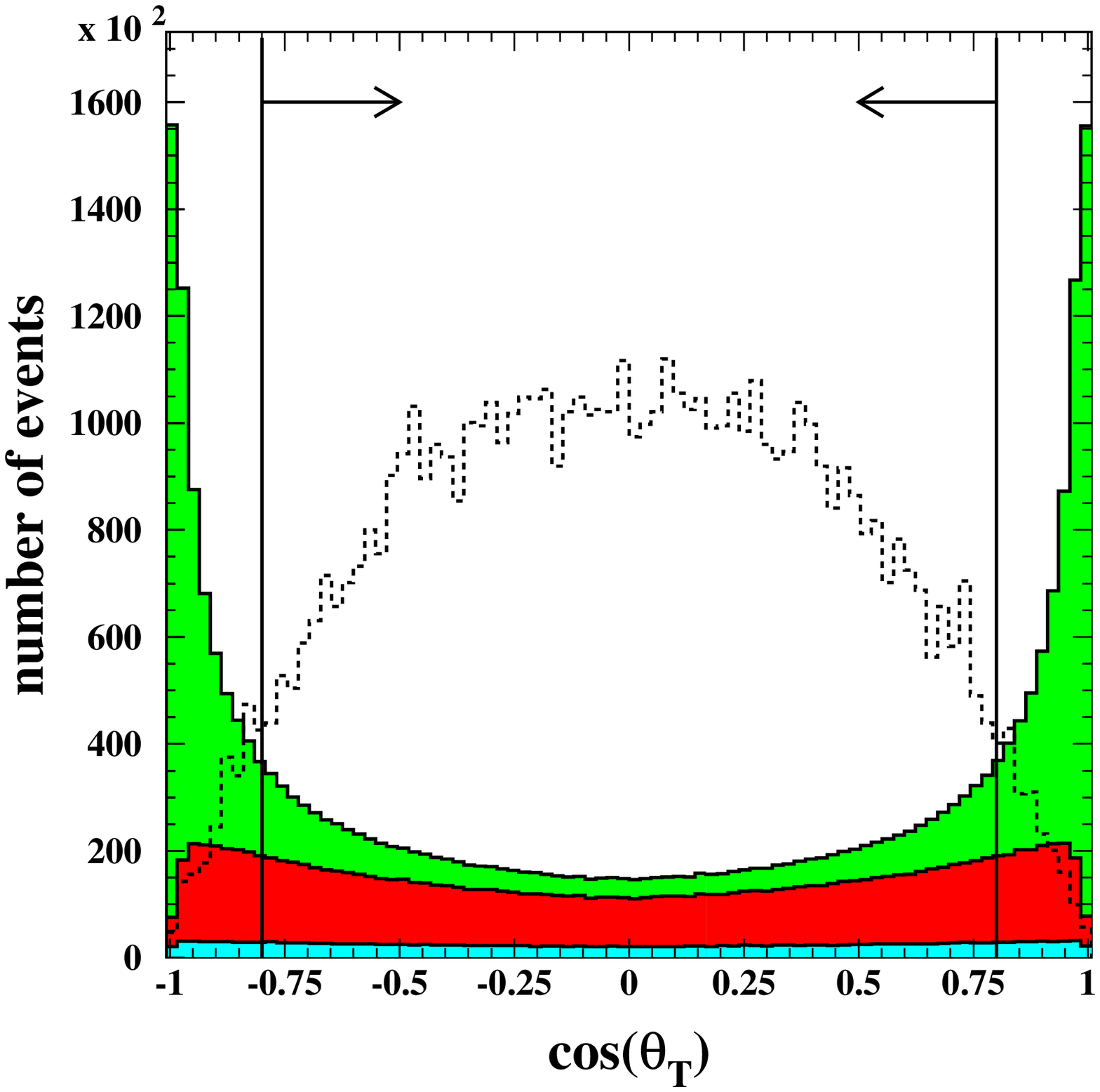}
\hspace{-4mm}
\includegraphics[width=0.5\textwidth]{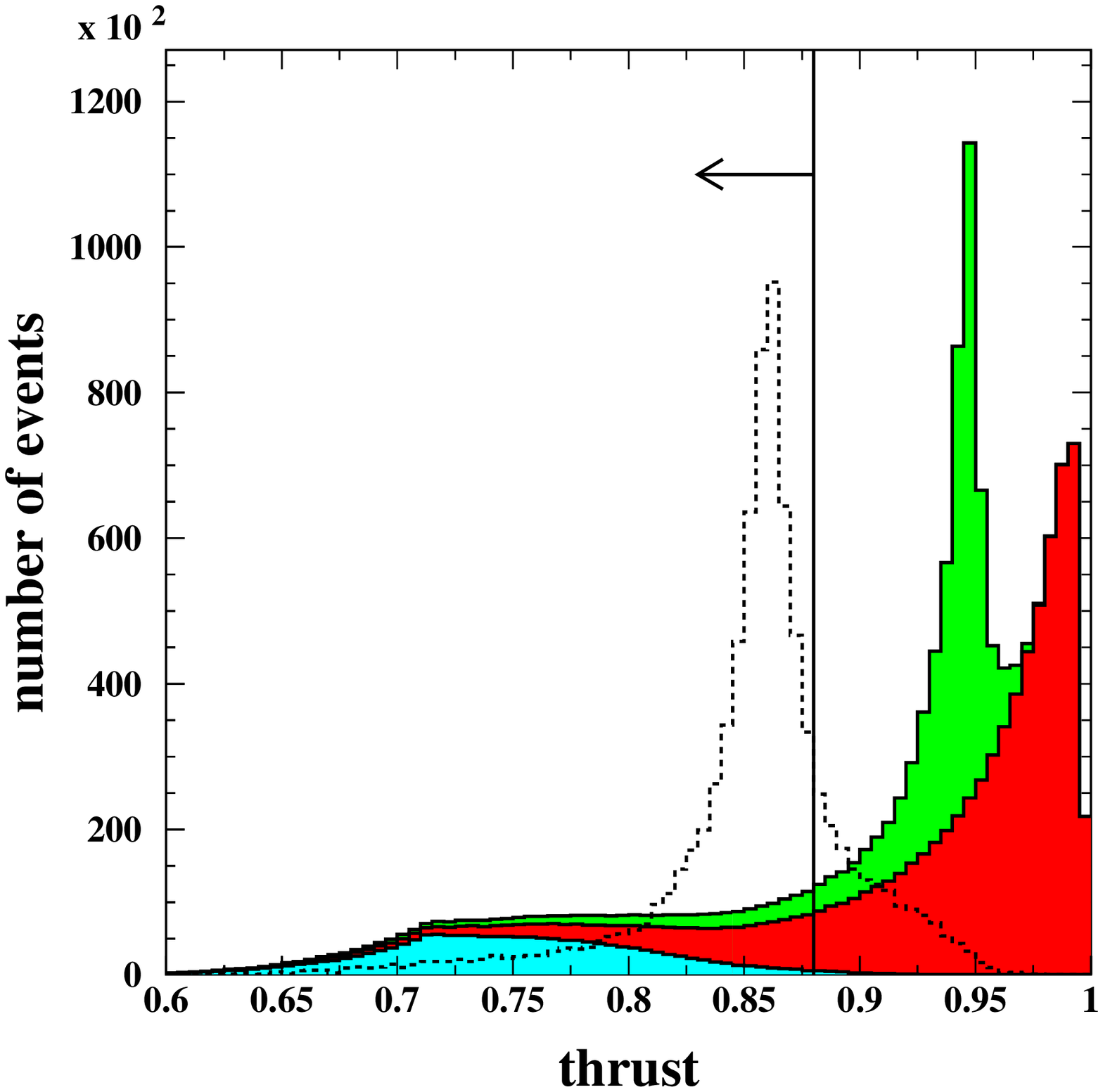}
\caption{Distributions of the selection variables (visible energy, number of tracks per jet $N_{tracks}/jet$, $\cos\theta_T$ and a thrust value) in the  
$\HAbbbb$ channel with ($\mH$,$\mA$) = (150,100) GeV at $\sqrts$ = 500 GeV.
The signal distributions are shown with arbitrary normalisation.
The vertical lines and arrows indicate cuts, imposed on these variables. The distributions are shown after all cuts preceding the current variable. 
}
\label{fig:var1_bbbb}
\end{center}
\end{figure}

\begin{figure}[p]
\begin{center}
\includegraphics[width=0.5\textwidth]{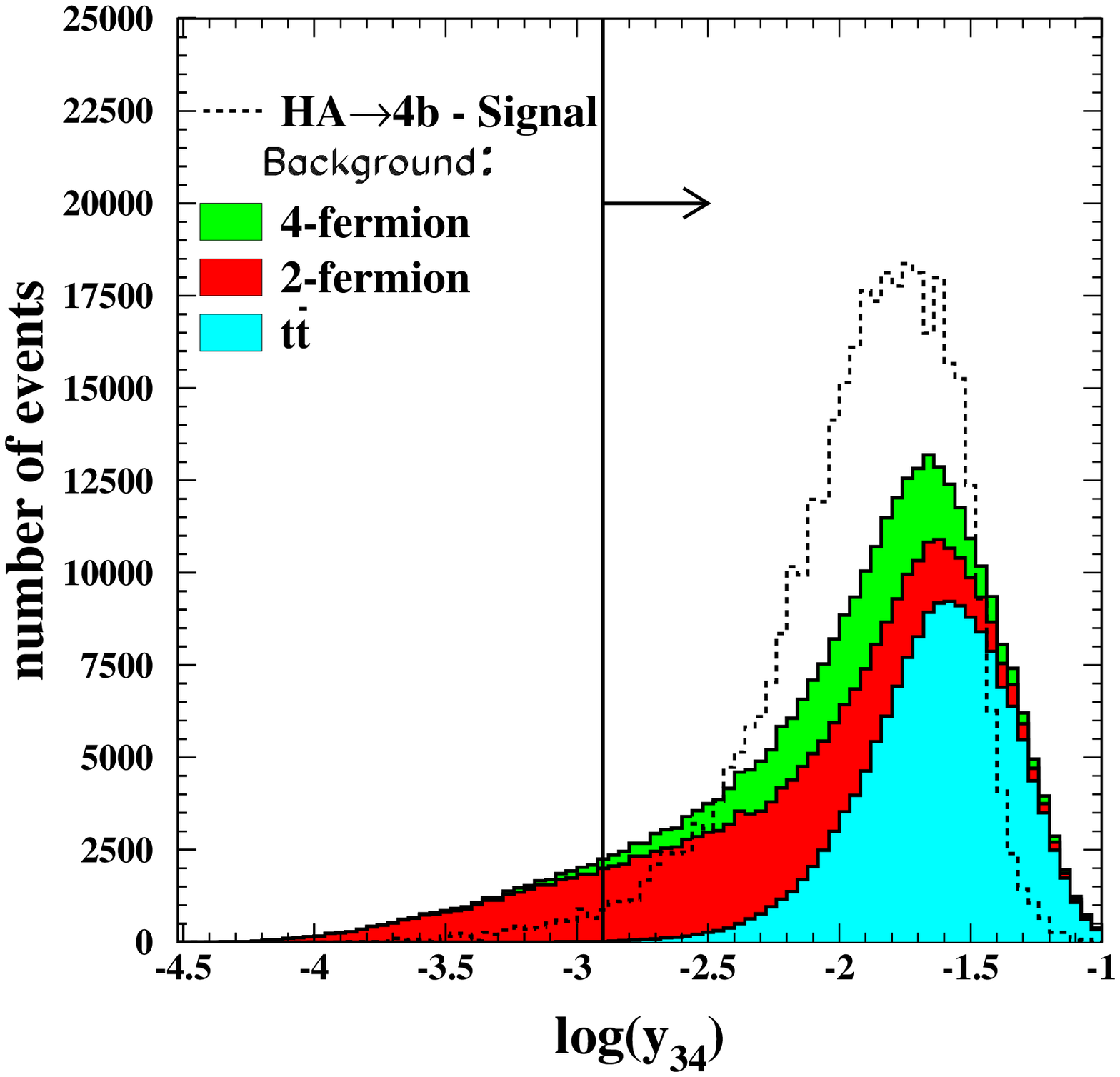}
\hspace{-4mm}
\includegraphics[width=0.5\textwidth]{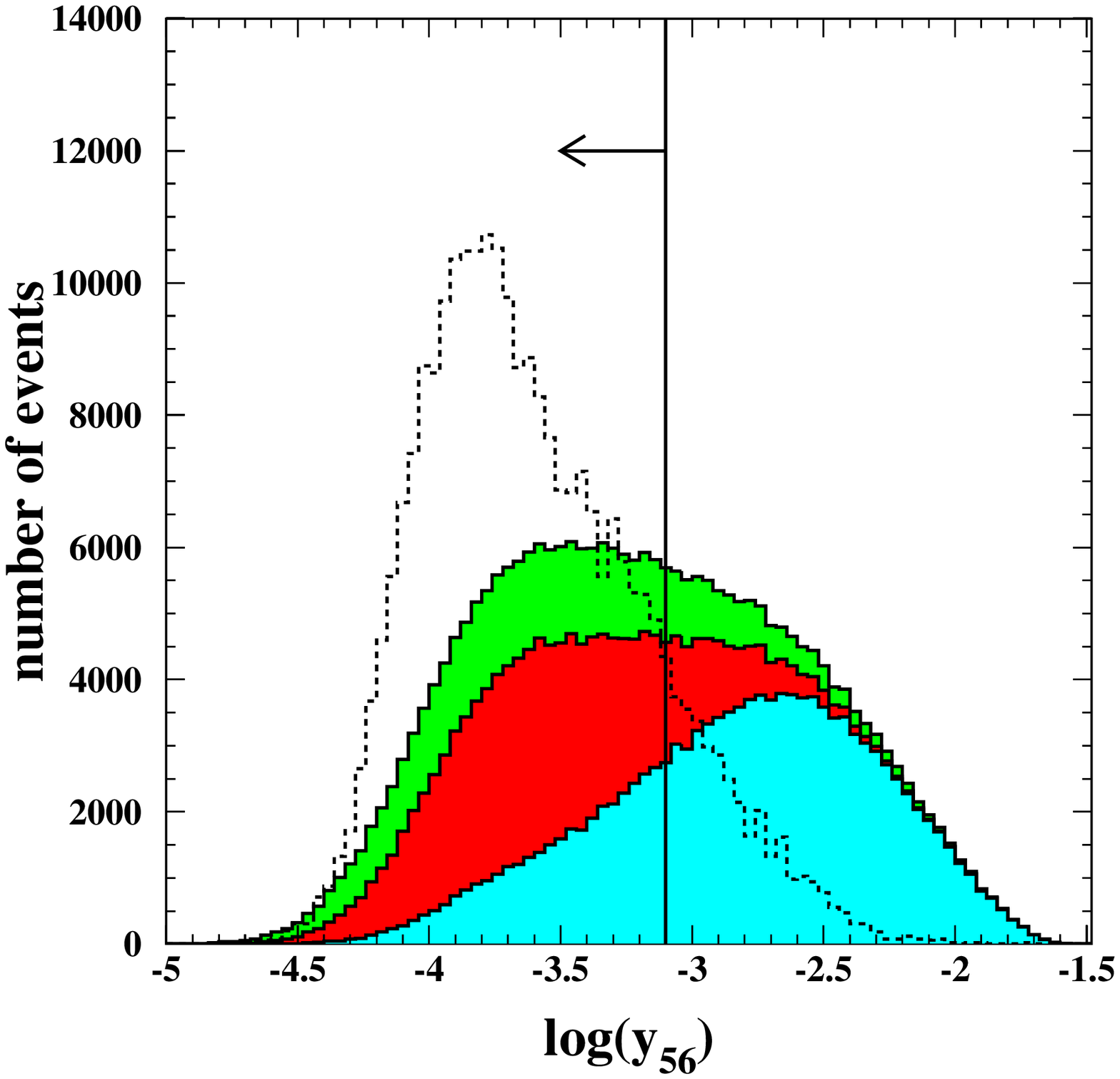}
\includegraphics[width=0.5\textwidth]{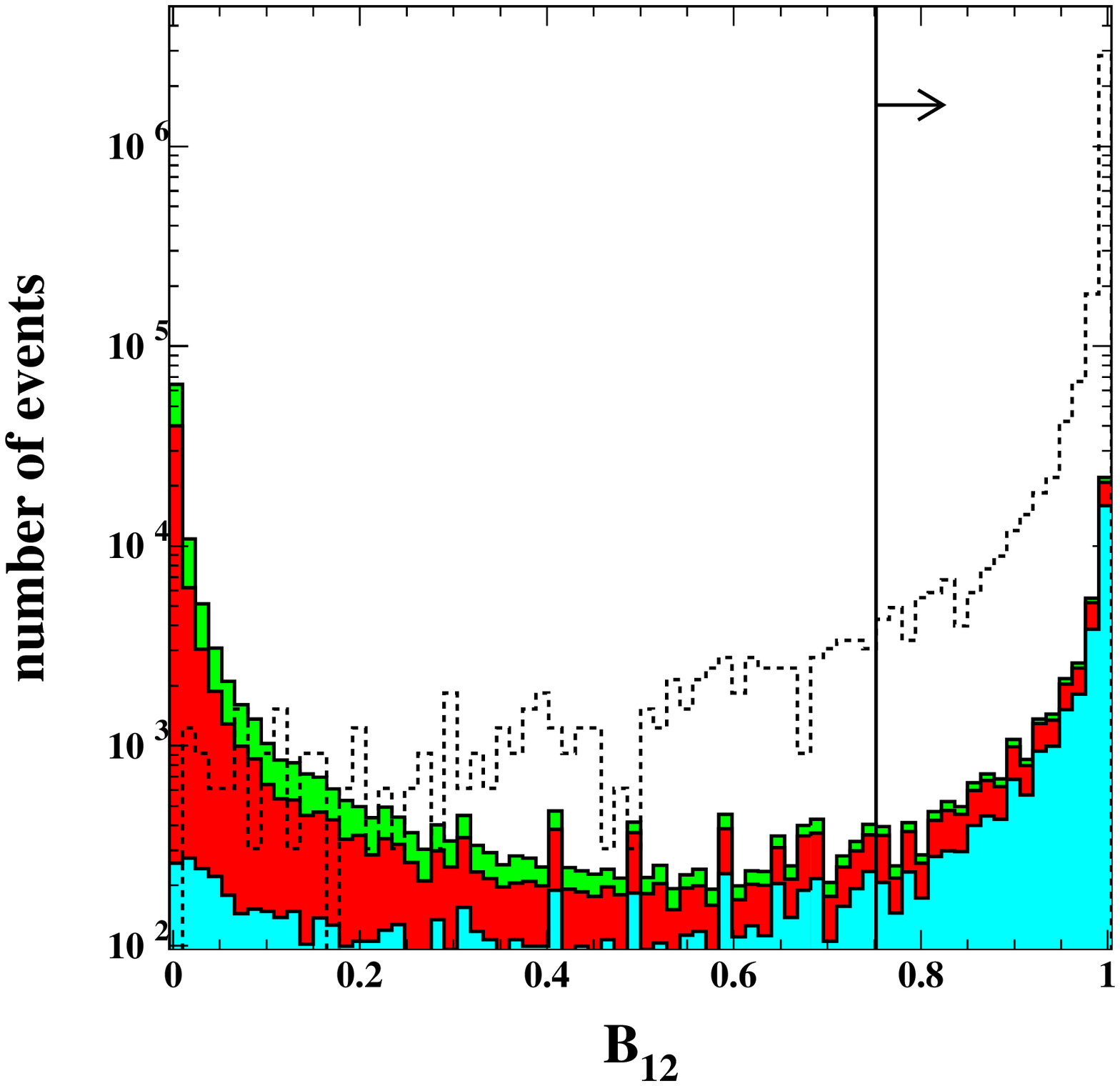}
\hspace{-4mm}
\includegraphics[width=0.5\textwidth]{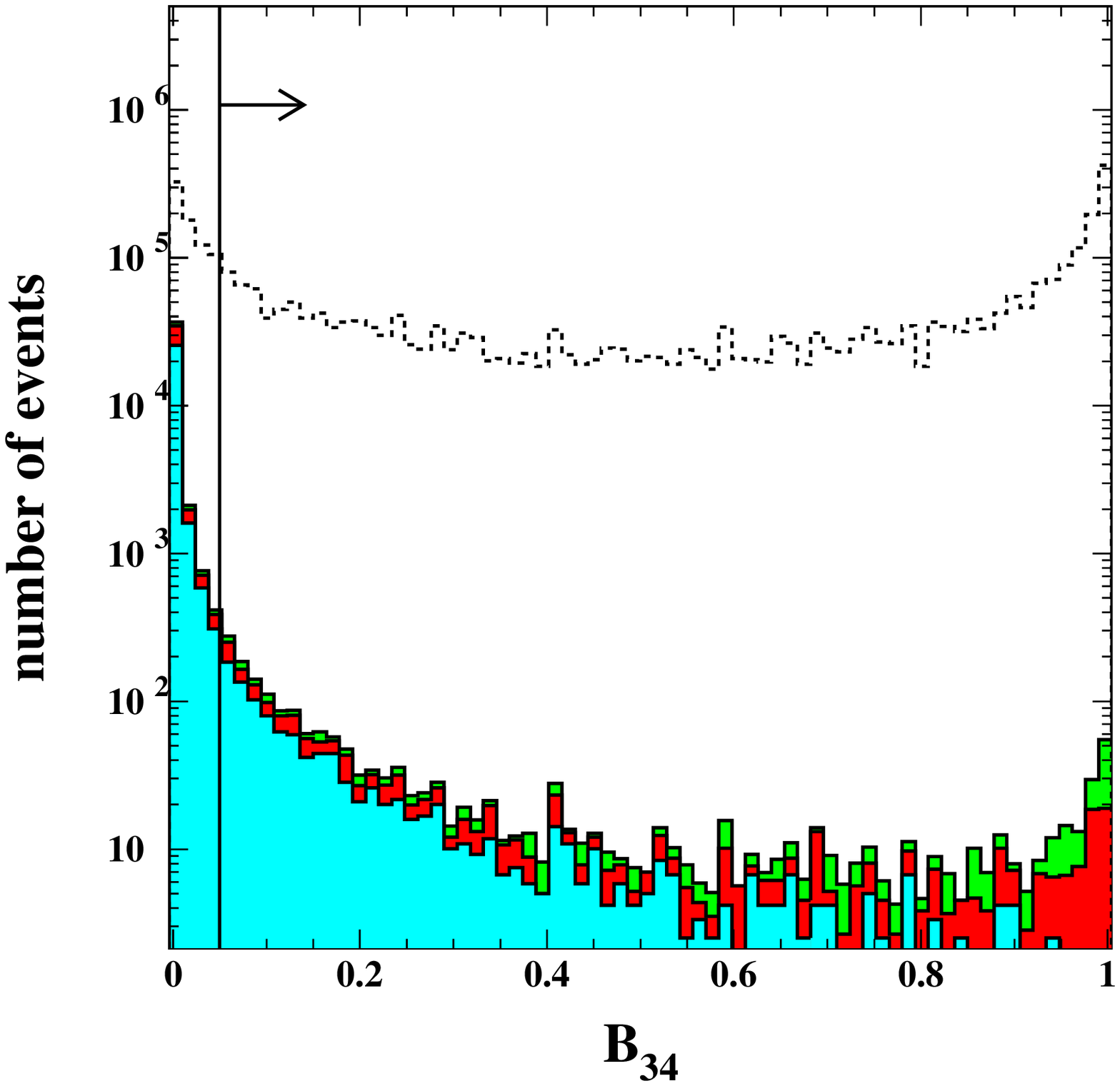}
\caption{Distributions of the selection variables (jet resolution parameters $\log_{10}{y_{34}}$ and $\log_{10}{y_{56}}$ and b-tag variables $B_{12}$ and $B_{34}$)  in the  $\HAbbbb$ 
channel with ($\mH$,$\mA$) = (150,100) GeV at $\sqrts$ = 500 GeV. The signal
distributions are shown with arbitrary normalisation. 
The vertical lines and arrows indicate cuts, imposed on these variables. The distributions are shown after all cuts preceding the current variable.
\label{fig:var2_bbbb}
}
\end{center}
\end{figure}

Events of the $\bbbb$ topology are characterized by four high
multiplicity hadronic jets, containing decay products 
of b-hadrons. A cut-based technique is employed to separate signal 
from background. Selection criteria are optimized separately for 
500 GeV and 800 GeV center-of-mass energies. Each event is required 
to pass the following cuts:
\begin{enumerate}
\item{Total energy deposited in the detector, the visible 
energy $\Evis$, must be greater than 340 GeV (600 GeV) 
for 500 GeV (800 GeV) center-of-mass energies.}
\item{Each event is forced into four jets using the DURHAM algorithm~\cite{DURHAM} and the number of tracks per jet is required to be greater than three.}
\item{To separate centrally produced H and A bosons from the 
WW and ZZ events, peaking in forward/backward direction, 
we apply a cut on the polar angle of the thrust vector~\cite{pythia}, $|\cos\theta_T|<0.8$.} 
\item{Further suppression of the WW and ZZ backgrounds is achieved by 
requiring the event thrust value to be less than 0.88.}
\item{Two fermion background is suppressed by applying a cut on
the \newline DURHAM jet resolution parameter, for which the event changes from four 
to three jets, $\log_{10}{y_{34}}\geq -2.9$. }
\item{High multiplicity six-jet events originating from the $\tt$  
production are reduced by requiring the number of enflow (energy flow) objects in the event to be less than 130. This cut is applied only at $\sqrts$ = 500 \GeV.}
\item{The background from $\tt$ events is further reduced by applying a 
cut on the jet resolution parameter, for which the event changes from 
six to five jets, $\log_{10}{y_{56}}$ $\leq$ -3.1 (-2.8) at 
$\sqrts$ = 500 GeV (800~\GeV).}
\item{Finally, we make use of the b-tag information to enhance the purity
of the selected event sample. First, the b-tag variable for each jet 
is calculated as described in Section 4. The four b-tag variables are sorted in descending order,
$B_1>B_2>B_3>B_4$. Two quantities $B_{12}$, $B_{34}$ are 
then defined as 
\begin{displaymath}
B_{12}=\frac{B_1B_2}{B_1B_2+(1-B_1)(1-B_2)},
\end{displaymath}
\begin{displaymath}
B_{34}=\frac{B_3B_4}{B_3B_4+(1-B_3)(1-B_4)}.
\end{displaymath}
The value of $B_{12}$ has to be greater than 0.75 (0.6) 
at $\sqrts$ = 500 GeV (800 GeV). The value of $B_{34}$ is required to be greater
than 0.05 independent of the center-of-mass energy.}
\end{enumerate}
The numbers of expected signal and background events, retained after selection, and signal efficiency for the example points with ($\mH$,$\mA$) = (150,100) GeV at $\sqrts$ = 500~\GeV and ($\mH$,$\mA$) = (300,250) at $\sqrts$ = 800~\GeV are presented in Tables~\ref{tab:cutflow_bbbb_500},~\ref{tab:cutflow_bbbb_800}. Fig.~\ref{fig:var1_bbbb} and~\ref{fig:var2_bbbb} show the distributions of 
the selection variables for the different sources of 
background and for the signal at $\sqrts$ = 500~\GeV. The cutflow table for this case shows that such cuts like those on visible energy, number of tracks per jet, polar angle on the thrust vector and b-tag cuts are very effective to separate signal from the background. Especially should be stressed that after all kinematic selection i.e. before b-tag cuts the signal efficiency is 54\% while the background is drastically reduced (1.6\% from the initial value). After b-tag cuts only 1.3\% of the background events, left after kinematic selection, remains, while the signal efficiency is 43\%. This confirms, that the new tool for b-tagging is very powerful. The signal efficiencies, number of signal events and total background for different Higgs boson masses after selection cuts are given in Table~\ref{tab:bbbb_effy}.    

Events accepted in the final sample are subjected to a 4C kinematic fit.
For each of the three possible di-jet pairings, the di-jet mass sum and 
the di-jet mass difference are reconstructed. For the example point the mass sum and the mass difference presented in Fig.~\ref{fig:kf_bbbb} show that due to the kinematic fit the mass reconstruction resolution can be improved significantly. 

%
%
\subsection{The $\HAbbttbb$ Channels}
\label{label:bbttcuts}
The signature of the $\bbtt$ final state consists of two high-multiplicity hadronic jets enriched with the decay
products of b-hadrons and two low-multiplicity jets initiated 
by $\tau$-leptons. 

Initially, a preselection of high-multiplicity events compatible
with the four-jet topology is applied. Events are required to fulfill
the following criteria.
\begin{enumerate}
\item{The leptonic and two photon events are rejected by applying
lower cut on the number of energy flow objects, $\Nefo$ $>$ 30.} 
\item{Genuine six-jet events, resulting from the $\tt$ 
production and characterized by very large particle multiplicity, 
are partially suppressed by an upper cut on the number of energy flow 
objects, $\Nefo$ $<$ 120.}
\item{The visible energy $\Evis$ must be greater than 
250 GeV (400 GeV) at $\sqrts$ = 500 GeV (800 GeV). This cut suppresses background processes
characterized by large missing energy: $\ZZqqnn$ events, radiative 
returns to the Z resonance with a photon escaping into the beam pipe and $\Wen$, $\Zee$ processes with electrons escaping undetected 
in the forward direction. An upper cut  $\Evis$ $<$ 500 GeV (760 GeV) at 
$\sqrts$ = 500 GeV (800 GeV) allows partial 
rejection of the fully hadronic $\WWqqqq$ and $\ZZqqqq$ events without 
significant loss in the signal efficiency.}
\item{The event thrust value has to be less than 0.95. This cut disentangles
more spheric HA events from the WW, ZZ and $\qq$ backgrounds in 
which the high boost of the final state bosons/quarks results in large values 
of the event thrust.}
\item{To separate centrally produced H and A bosons from the WW 
and ZZ processes in which weak bosons tend to peak in the forward/backward 
directions, a cut on the polar angle of the thrust vector, 
$|\cos\theta_T|$ $<$ 0.9, is applied.}
\item{The event is discarded if it contains an $\ee$ or $\mumu$ pair
with invariant mass compatible with the Z mass within 5 GeV. 
This cut suppresses the $\ZZqqll$ background.}
\end{enumerate}
 
Events satisfying these criteria are resolved into four-jet topology
and $\tau$-lepton identification is performed. Two identification 
categories for $\tau$-leptons are introduced. A low multiplicity jet is assigned
to the first category if it fulfills one of the following tight conditions: 
\begin{itemize}
\item{A jet contains only one charged track, 
identified as an electron or muon. No neutral clusters are assigned to the jet.}
\item{A jet contains one charged track, identified 
as hadron, and not more than four neutral clusters
(one-prong hadronic decay of $\tau$-lepton).}
\item{A jet contains three charged tracks with unit total charge and 
has not more than two neutral calorimetric clusters (three-prong 
hadronic decay of $\tau$-lepton).}
\end{itemize}
To enhance the acceptance of signal events, the second identification 
category, imposing looser criteria, is introduced. According to these
criteria, a jet is identified as $\tau$-lepton if it has not more than two 
charged tracks and less than five neutral clusters. 

An event is accepted if at least one of the low multiplicity jets satisfies the tight identification criteria. 
Signal efficiencies and numbers of signal and background 
events remaining after preselection are given in 
Table~\ref{tab:sgnl_pres_bbtt}.

\begin{figure}
\begin{center}
\includegraphics[width=0.5\textwidth]{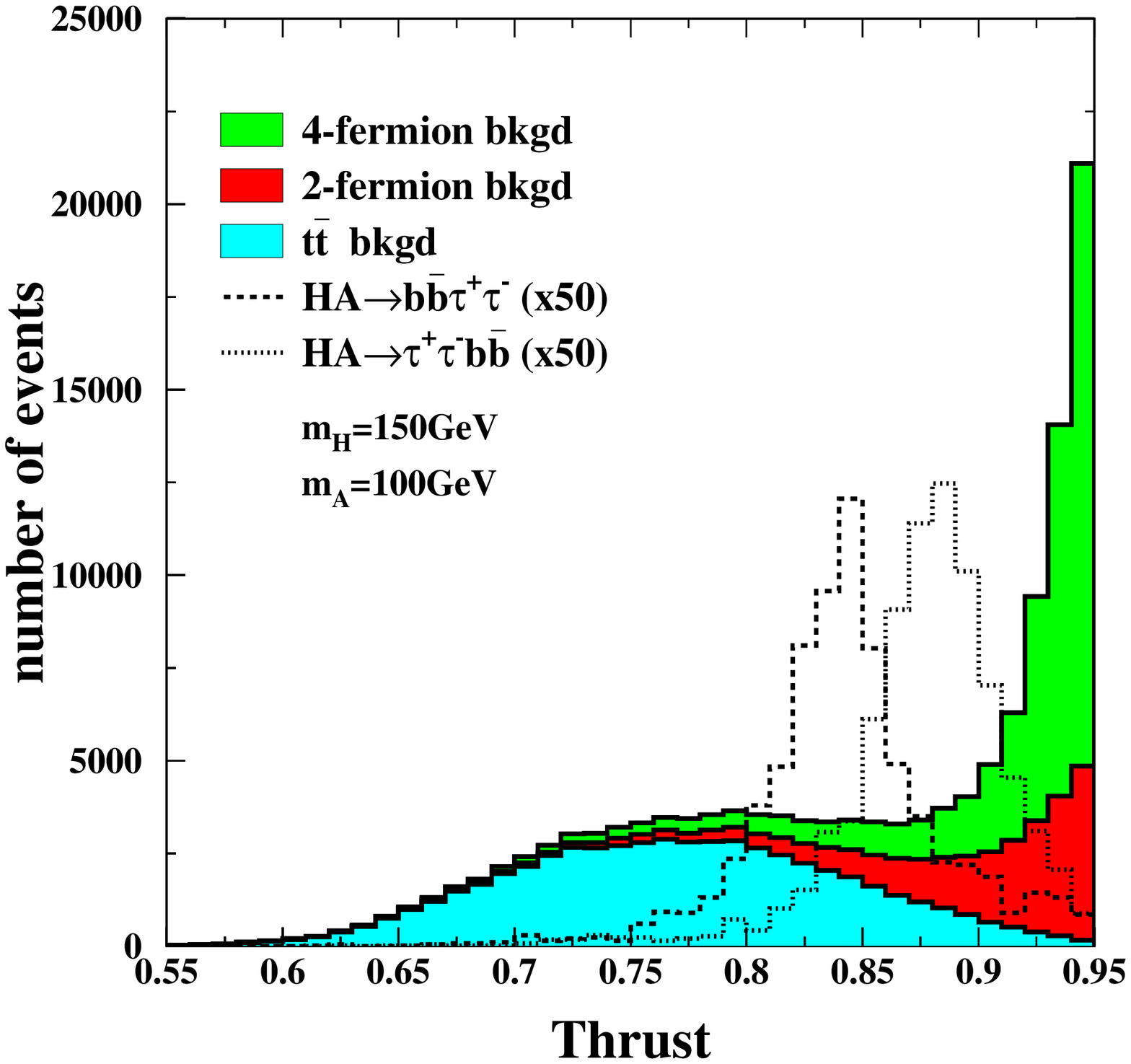}
\hspace{-4mm}
\includegraphics[width=0.5\textwidth]{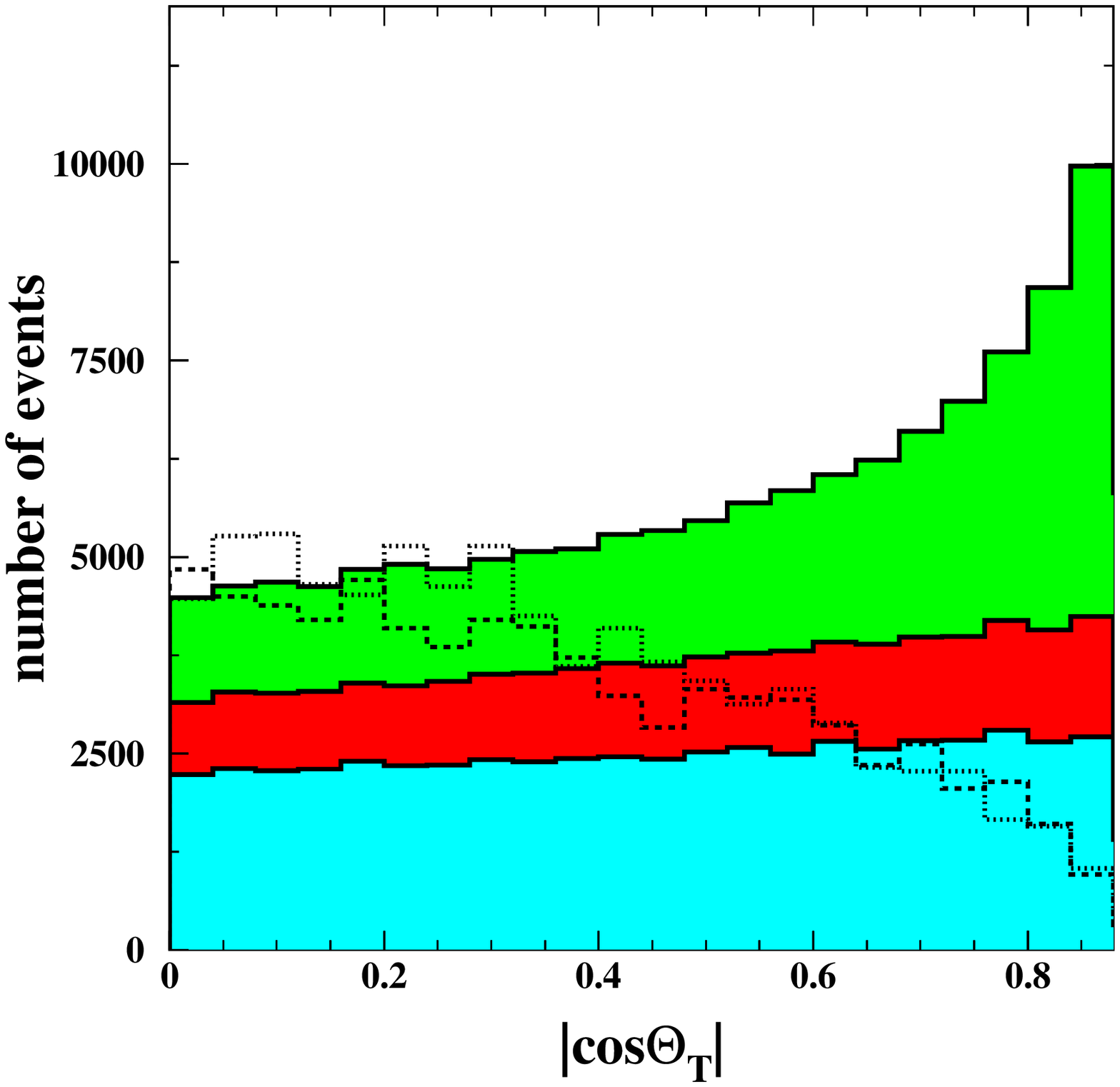}
\includegraphics[width=0.5\textwidth]{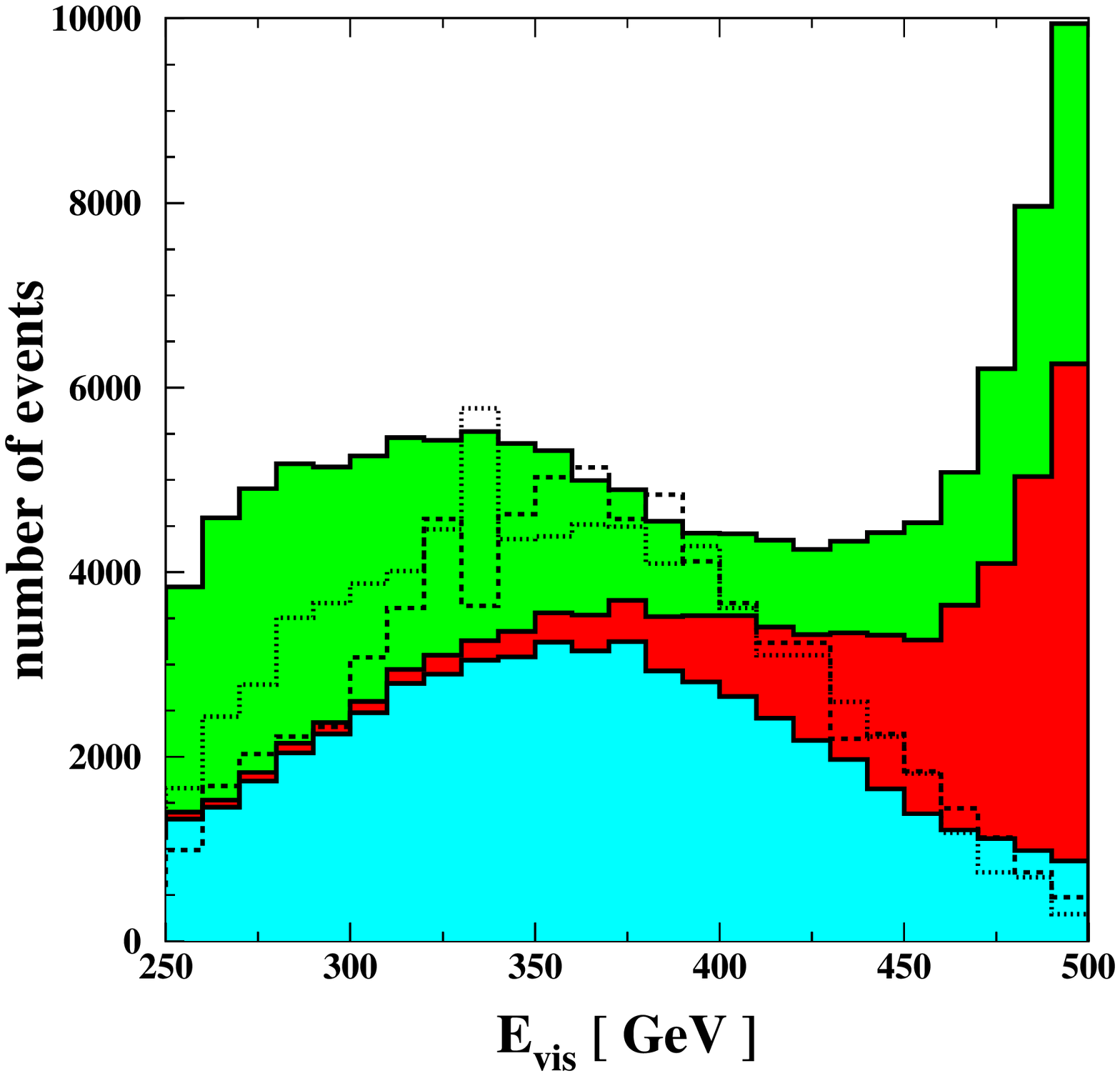}
\hspace{-4mm}
\includegraphics[width=0.5\textwidth]{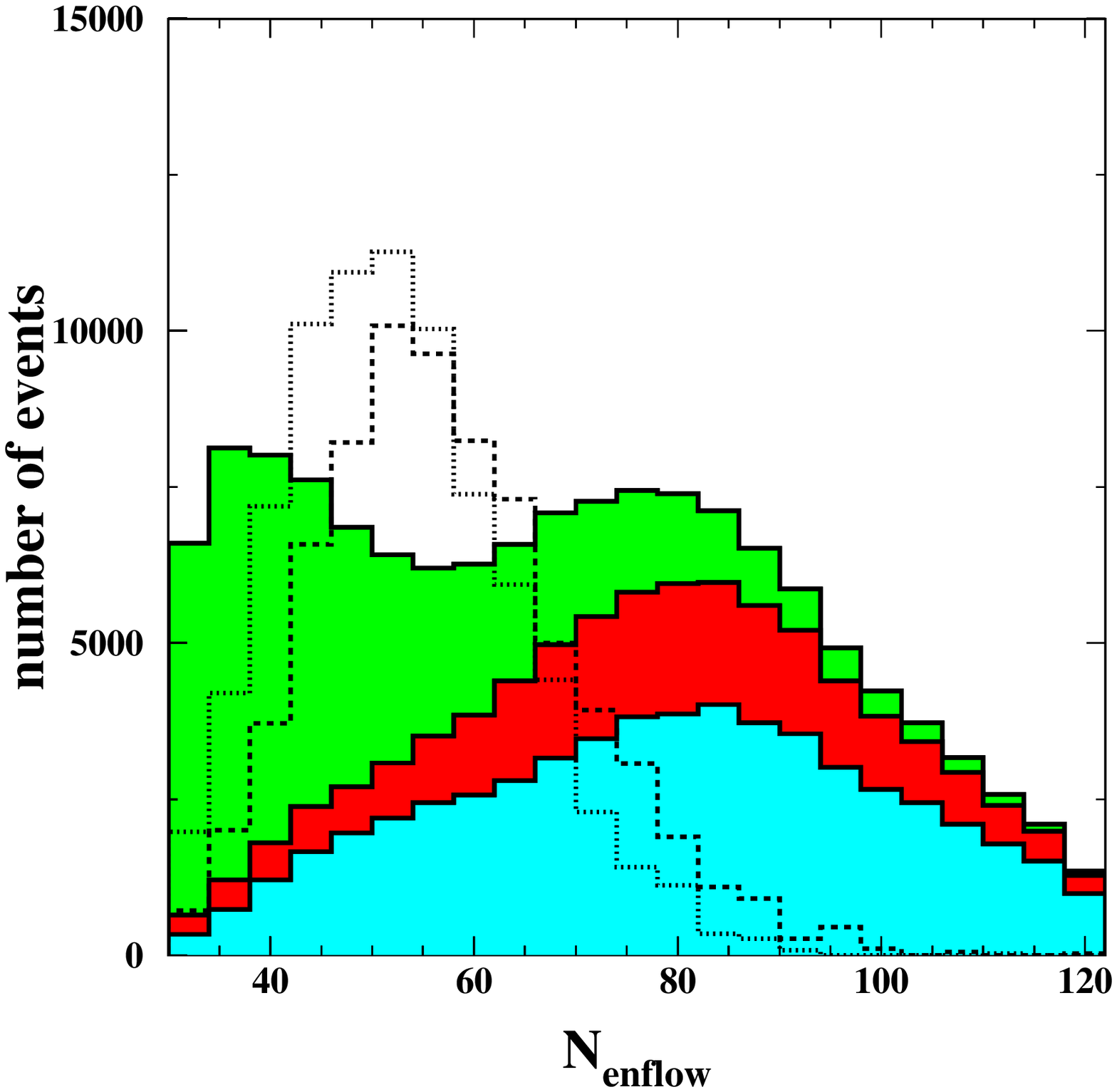}
\end{center}
\caption{Variables used to construct signal likelihood (thrust value, $\cos\theta_T$, visible energy $E_{vis}$ and number of energy flow objects $N_{enflow}$) in the $\HAbbttbb$ channels with ($\mH$,$\mA$) = (150,100) GeV  at $\sqrts$ = 500 GeV. The distributions are shown after cuts 1-6 (Section~\ref{label:bbttcuts}).\label{fig:var1_bbtt}
}
\end{figure}

\begin{figure}
\begin{center}
\includegraphics[width=0.5\textwidth]{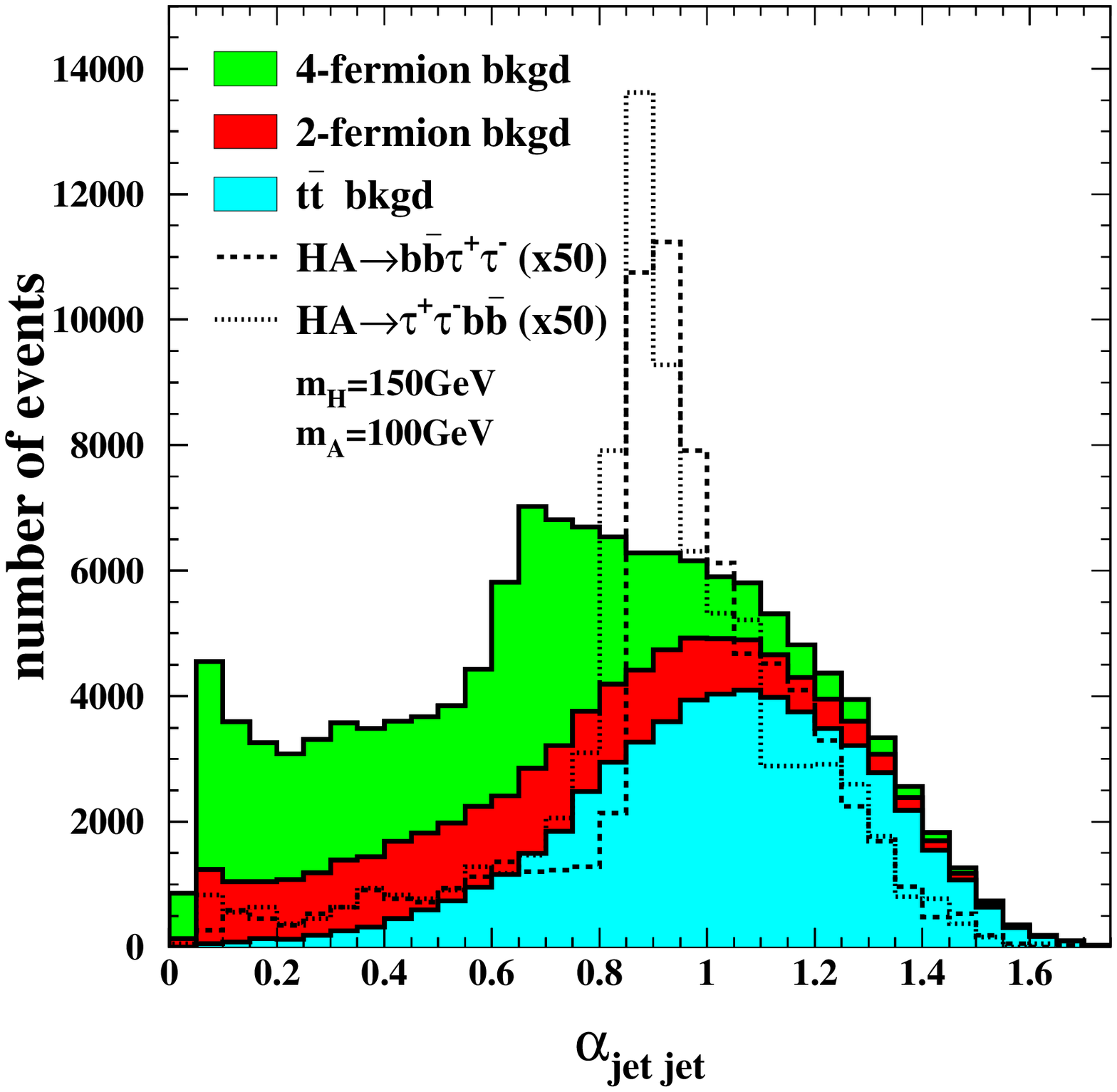}
\includegraphics[width=0.5\textwidth]{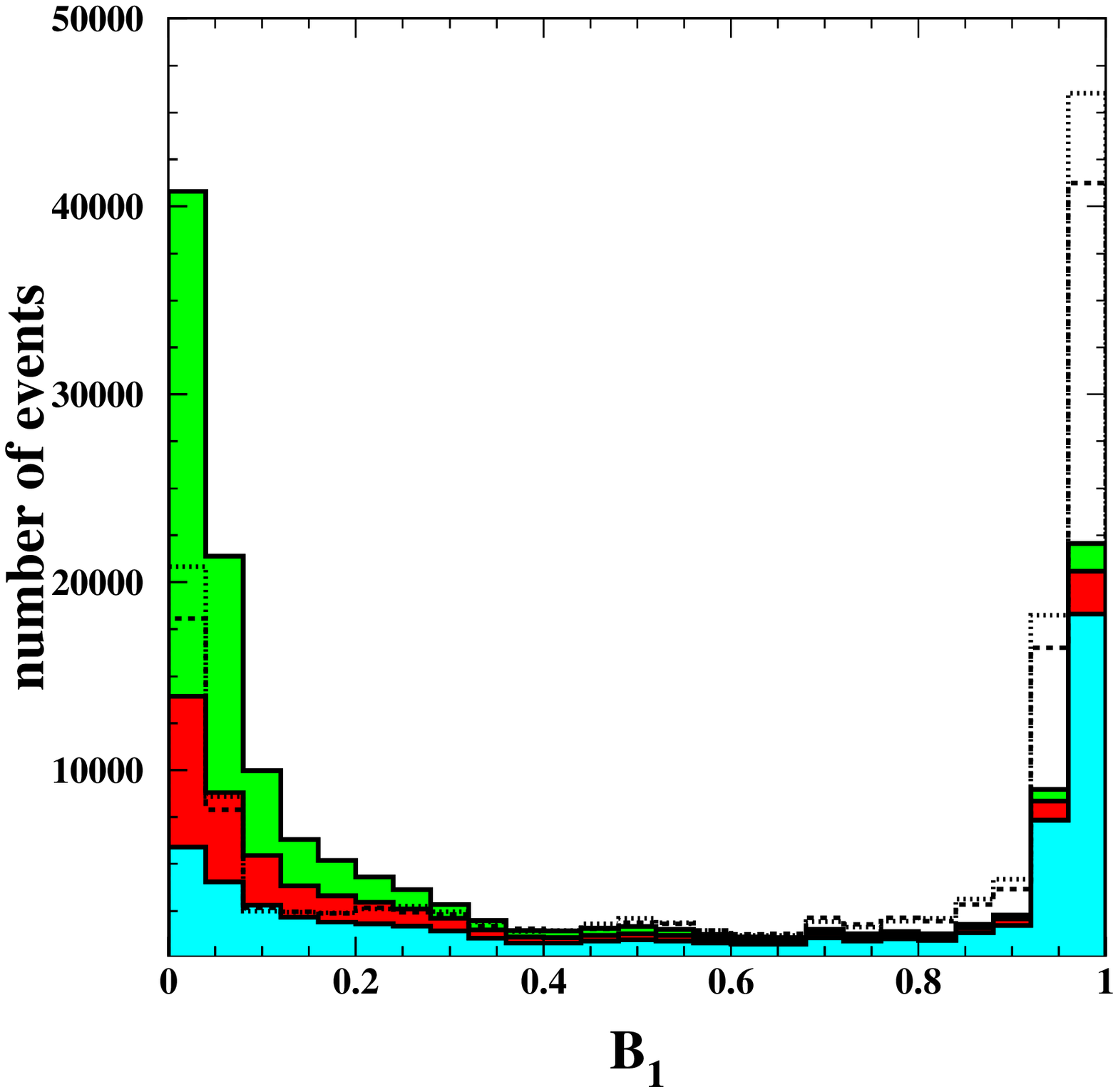}
\hspace{-4mm}
\includegraphics[width=0.5\textwidth]{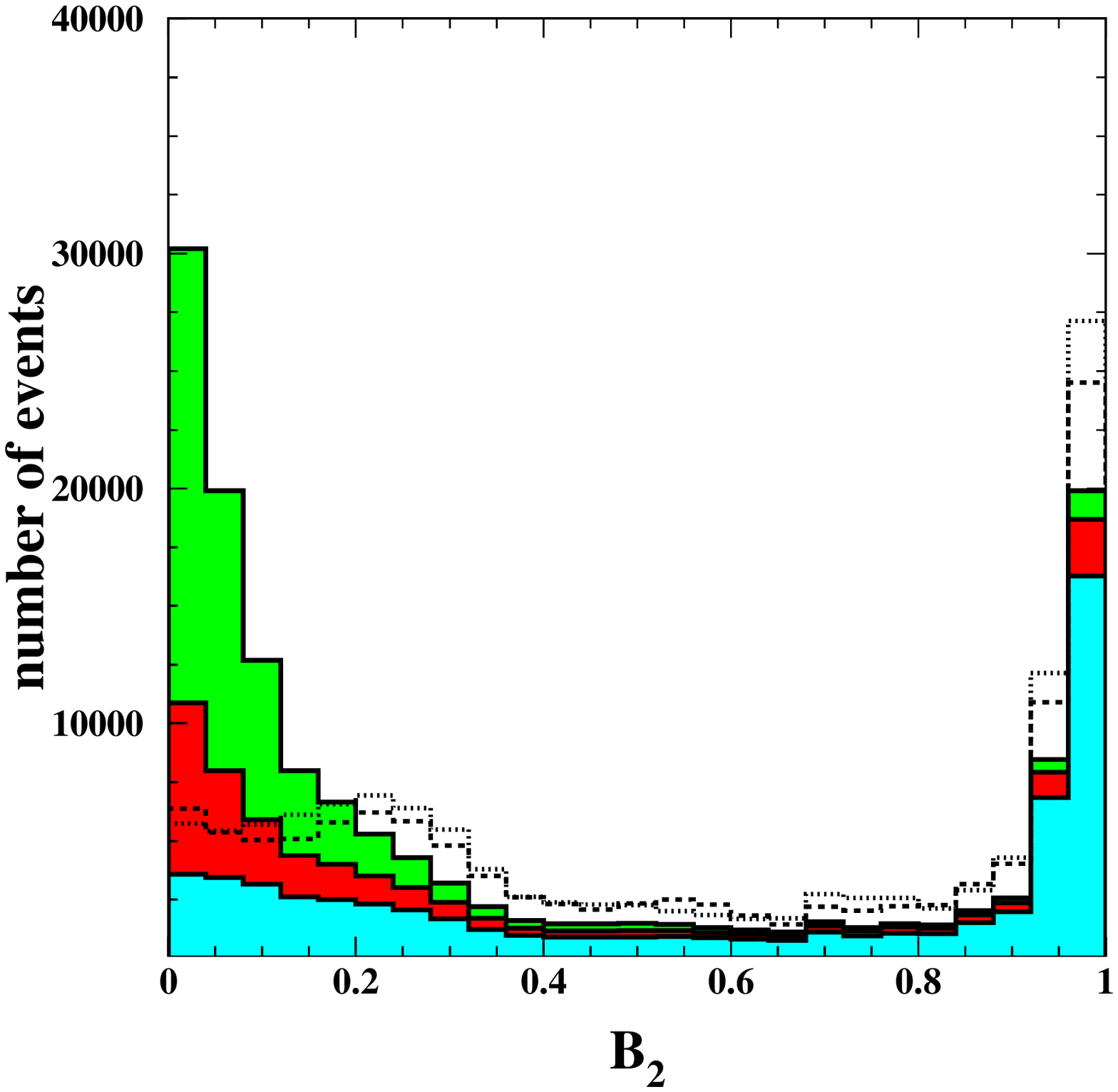}
\end{center}
\caption{Variables used to construct signal likelihood (minimal opening angle between any two jets $\alpha_{jet jet}$ and b-tag variables $B_{1}$ and $B_{2}$) in the $\HAbbttbb$ channels with ($\mH$,$\mA$) = (150,100) GeV  at $\sqrts$ = 500 GeV. The distributions are shown after cuts 1-6 (Section~\ref{label:bbttcuts}).
\label{fig:var2_bbtt}
}
\end{figure}

\begin{figure}
\begin{center}
\includegraphics*[width=1.1\textwidth]{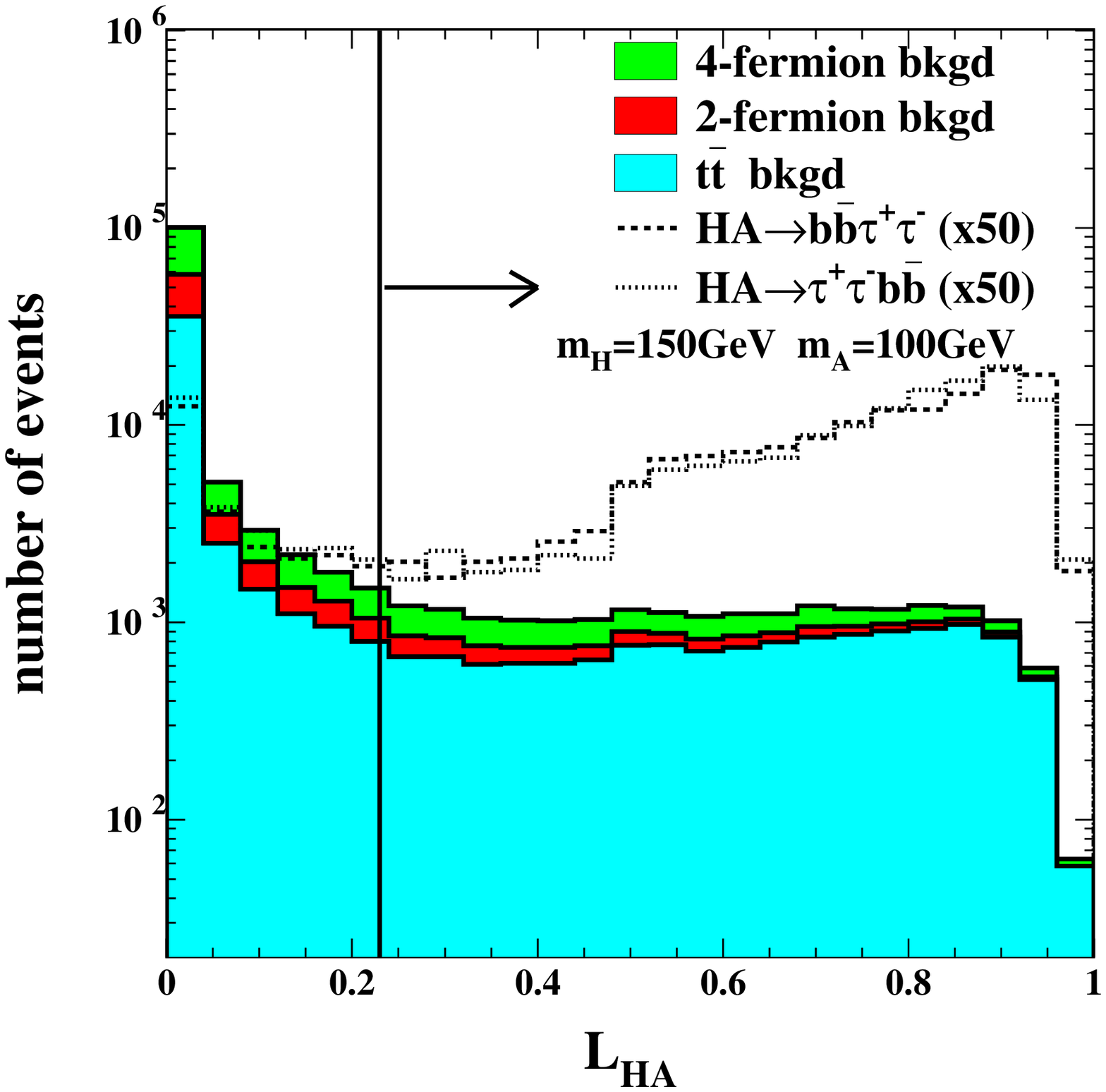}
\end{center}
\caption{Distribution of $\LHA$ in the $\HAbbttbb$ channels with
($\mH$,$\mA$) = (150,100) GeV at $\sqrts$ = 500 GeV.
The vertical line and arrow indicate cut placed on the likelihood.
\label{fig:like_bbtt}
}
\end{figure}

In the next step, events are selected into the final sample
by means of a binned likelihood technique. The signal likelihood, $\LHA$
is built from the following variables:
\begin{itemize}
\item{event thrust, T;}
\item{the polar angle of the thrust vector, $\cos\theta_T$;}
\item{minimal opening angle between any two of the four jets;}
\item{b-tag variables of the two hadronic jets, $\Bh$ and $\Bl$
(hadronic jets are ordered by their energy);}
\item{missing energy;}
\item{number of energy flow objects, $\Nefo$.}
\end{itemize}
Fig.~\ref{fig:var1_bbtt} and~\ref{fig:var2_bbtt} show
the distributions of these variables. The distribution of 
$\LHA$ is shown in Fig.~\ref{fig:like_bbtt} for 
($\mH$,$\mA$) = (150,100) GeV at $\sqrts$ = 500 GeV. An event is accepted in the 
final sample if the value of $\LHA$ is greater than certain 
threshold optimized separately for each Higgs boson
mass hypothesis considered to yield maximal significance of the
signal. The number of signal  and background events and the 
signal efficiencies after final selection are given in 
Table~\ref{tab:sgnl_final_bbtt}. Signal efficiency ranges between 30 and 
45\%, and the number of background events contributing 
to the final sample between 2000 and 6000, depending on the Higgs mass.  

For events accepted in the final sample,
the di-jet and di-tau invariant masses are reconstructed
exploiting the 2C kinematic fit, described in Section 4.


\section{Results}
In the final step of the analysis, the spectra 
of the di-jet mass sum and difference, obtained in the $\HAbbbb$ channel,
and distributions of the di-jet and di-tau masses, 
reconstructed in the $\HAbbttbb$ channels, are used to determine 
Higgs boson properties. 

\subsection{Cross section and Mass}

\begin{figure}
\begin{center}
\includegraphics*[width=0.7\textwidth]{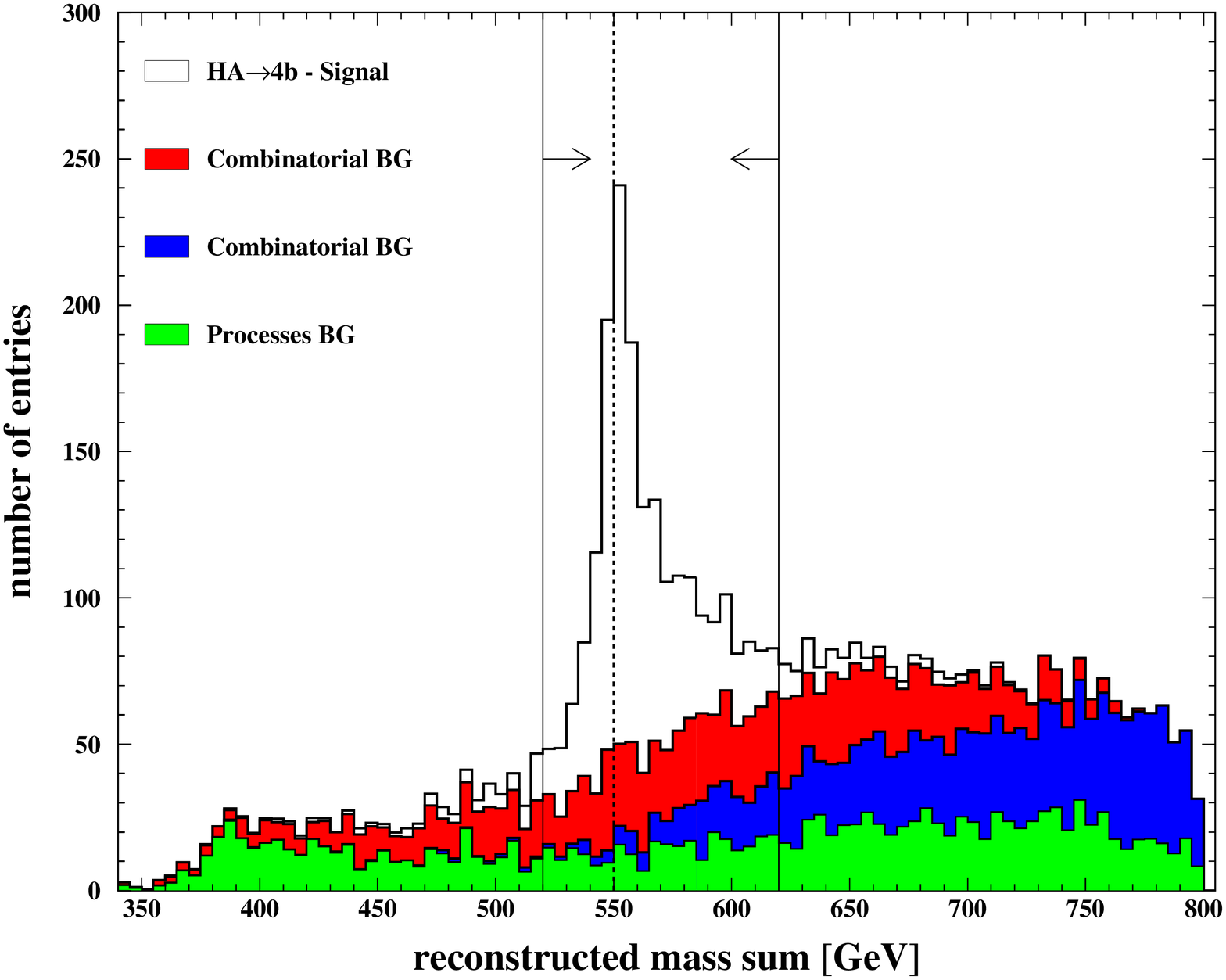}
\hspace{-8mm}
\includegraphics*[width=0.7\textwidth]{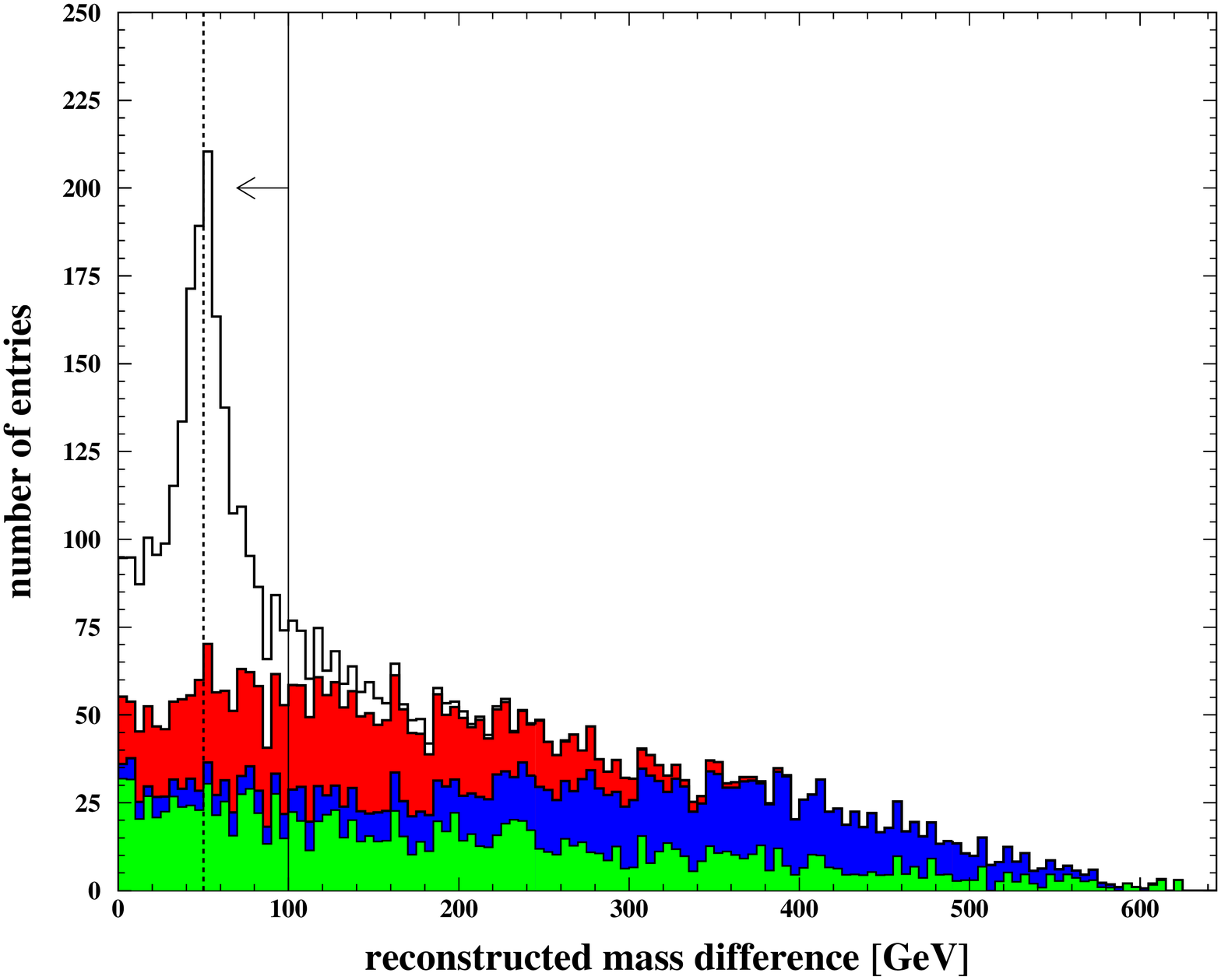}
\end{center}
\caption{
Distributions of the di-jet mass sum (upper figure) and di-jet mass difference (lower figure)
in the $\HAbbbb$ channel for Higgs boson mass hypothesis 
($\mH$,$\mA$) = (300,250) GeV at $\sqrts$ = 800 GeV after selection cuts and kinematic fit. The three components of the signal are shown separately: two combinatorial parts and one of the real signal. The background from the other processes is presented as well. 
\label{fig:combg}
}
\end{figure}

\begin{figure}
\begin{center}
\includegraphics*[width=0.7\textwidth]{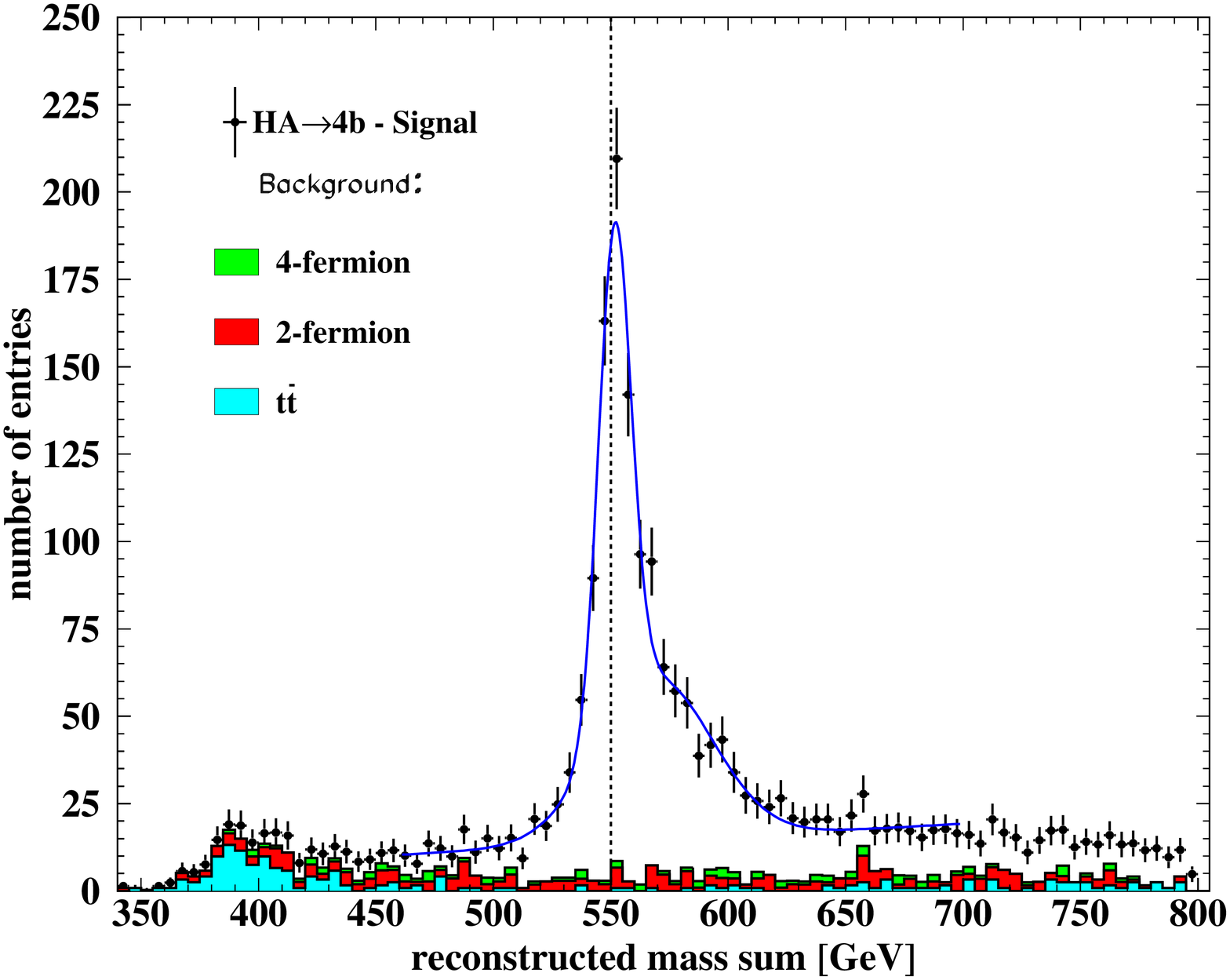}
\hspace{-8mm}
\includegraphics*[width=0.7\textwidth]{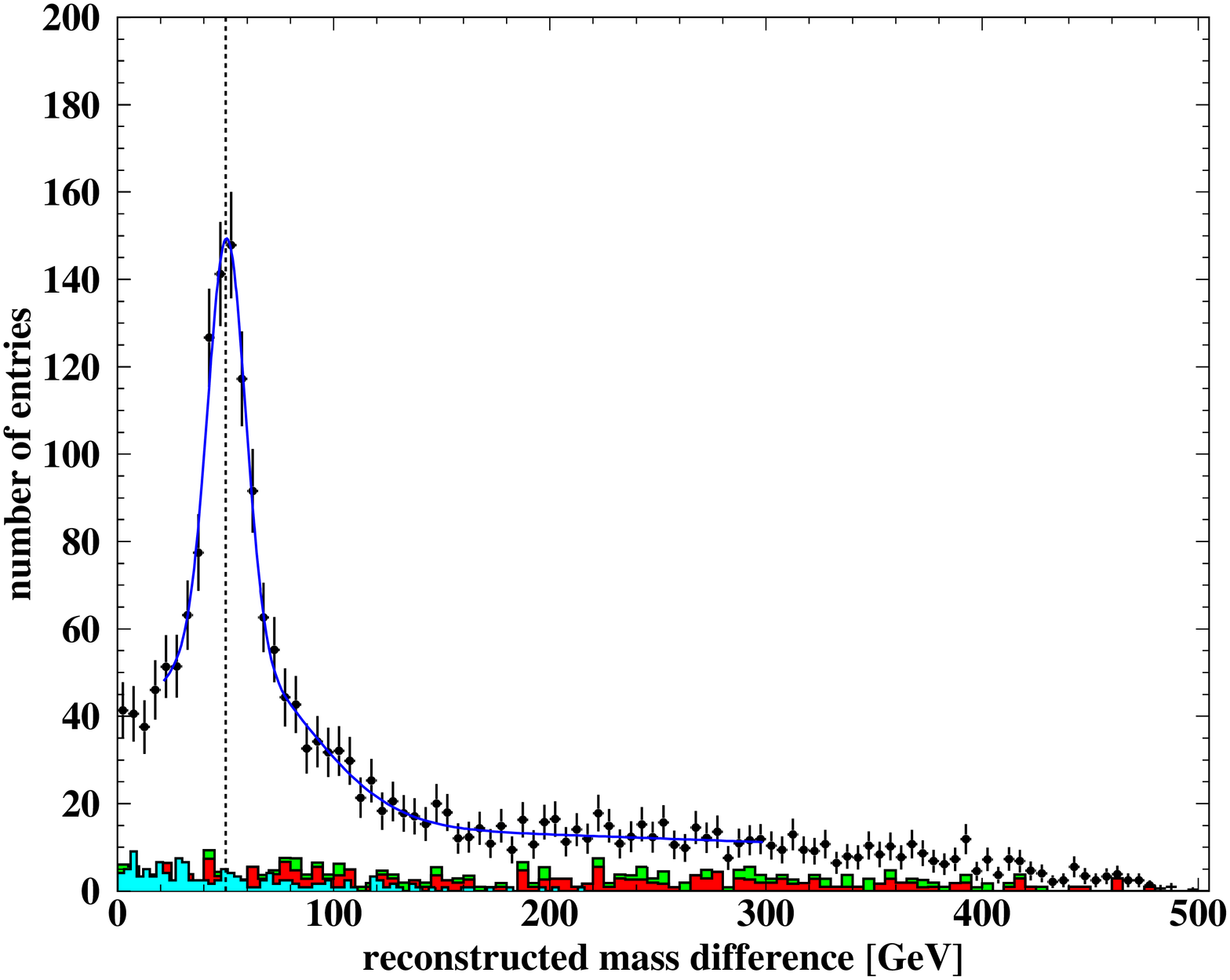}
\end{center}
\caption{
Upper figure: distribution of the di-jet mass sum after selection cuts, kinematic fit and cut on di-jet mass difference. Lower figure: distribution of the di-jet mass difference
after selection cuts, kinematic fit and cut on di-jet mass sum. Both distributions are in the $\HAbbbb$ channel for the Higgs boson mass hypothesis 
($\mH$,$\mA$) = (300,250) GeV at $\sqrt s$ = 800 GeV.
\label{fig:mass_bbbb}
}
\end{figure}

First, the analysis is performed assuming that the natural widths 
of the Higgs bosons are small compared to the detector
resolution.
As an example Fig.~\ref{fig:combg} shows the distributions of the di-jet mass sum and di-jet mass difference obtained after selection cuts and kinematic fit in the $\HAbbbb$ 
channel for the Higgs boson mass hypothesis of 
($\mH$,$\mA$) = (300,250) GeV at $\sqrts$ = 800 GeV. Three entries per event contribute to these distributions, corresponding to three possibilities to  
pair jets in the four-jet events. Two entries form a so-called combinatorial background. Fig.~\ref{fig:mass_bbbb} demonstrates the final di-jet mass sum and di-jet mass difference after the cut on the di-jet mass difference sum, respectively, as indicated by arrows in Fig.~\ref{fig:combg}. The signal efficiencies, number of signal events and total background for different Higgs boson masses after cuts on di-jet mass sum and difference are given in Table~\ref{tab:bbbb_effy}. The mass distributions are fitted separately
with the signal normalization, the sum and the difference 
of the Higgs boson masses as free parameters. The shapes of the 
signal distributions are parametrised using high statistics signal samples. 
Background is approximated with a polynomial function. From the fit, 
the sum and the difference of the Higgs boson masses
and errors on these quantities are determined.
The di-jet mass sum and difference are found to be weakly correlated 
quantities. Hence, the statistical errors on the masses of the 
Higgs bosons can be calculated as follows:
\begin{displaymath}
\delta \mH = \delta \mA = \frac{1}{2}
\sqrt{(\delta \Sigma^2 + \delta \Delta^2)},
\end{displaymath}
where $\delta \Sigma$ and $\delta \Delta$ are statistical
uncertainties in determination of the Higgs boson mass sum and difference.
The error on the topological cross section is calculated as
\begin{displaymath}
\delta\sigma/\sigma = \sqrt{N_S+N_B}/N_S.
\end{displaymath}
The notations $N_S$ and $N_B$ stand for 
the number of background and signal entries
within the windows in the di-jet mass sum and di-jet mass 
difference distributions, where the signal is accumulated. The boundaries
of these windows are optimized to yield a minimal relative error on 
the topological cross section.

\begin{figure}
\begin{center}
\includegraphics*[width=0.7\textwidth]{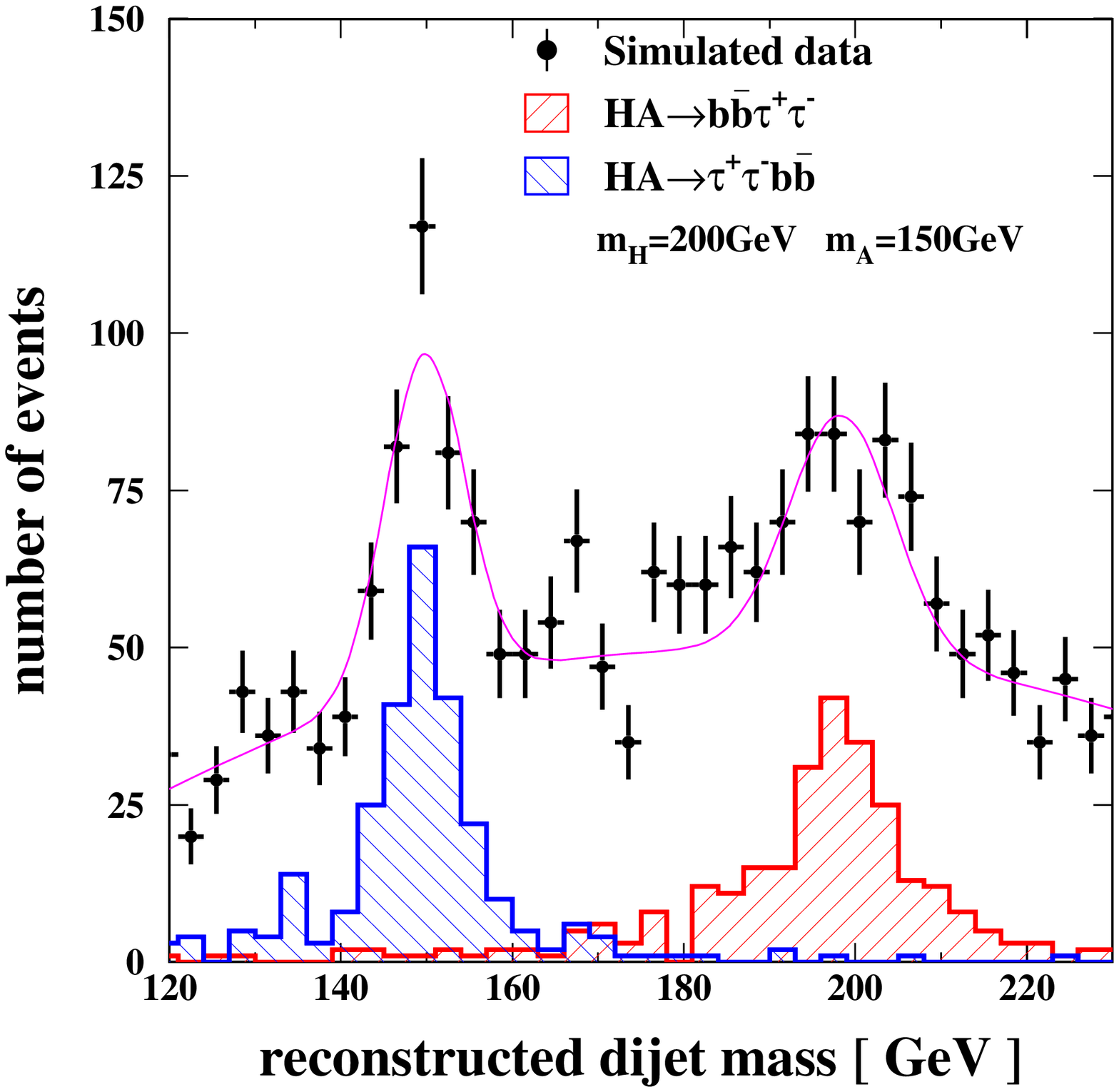}\\
\vspace{-10mm}
\includegraphics*[width=0.7\textwidth]{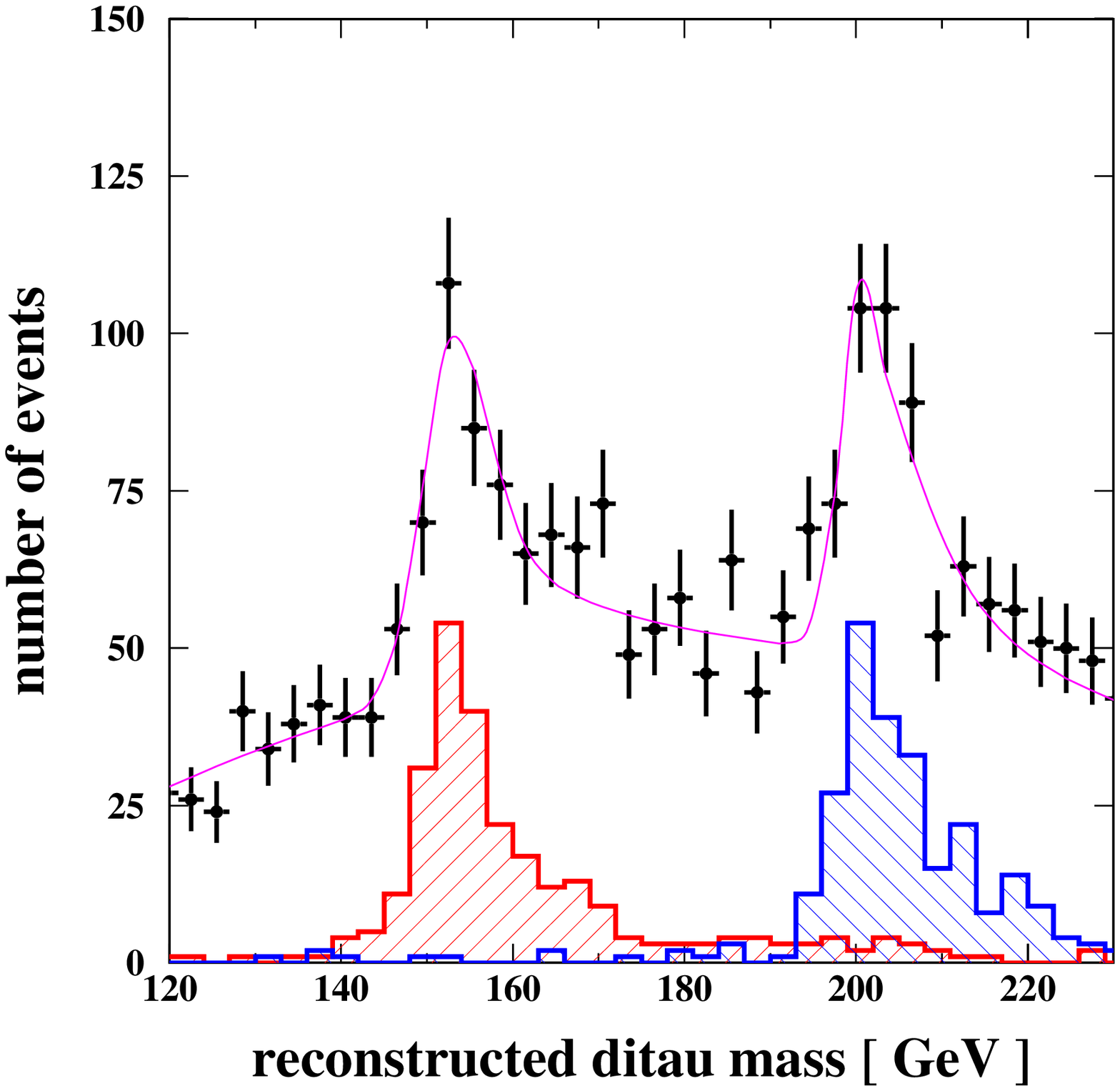}
\end{center}
\caption{Distributions of di-jet mass and di-tau mass in the $\HAbbttbb$ channels for the case of ($\mH$,$\mA$) = (200,150) GeV at $\sqrt s$ = 500 GeV.
\label{fig:mass_bbtt}
}
\end{figure}

\begin{figure}
\begin{center}
\includegraphics*[width=0.7\textwidth]{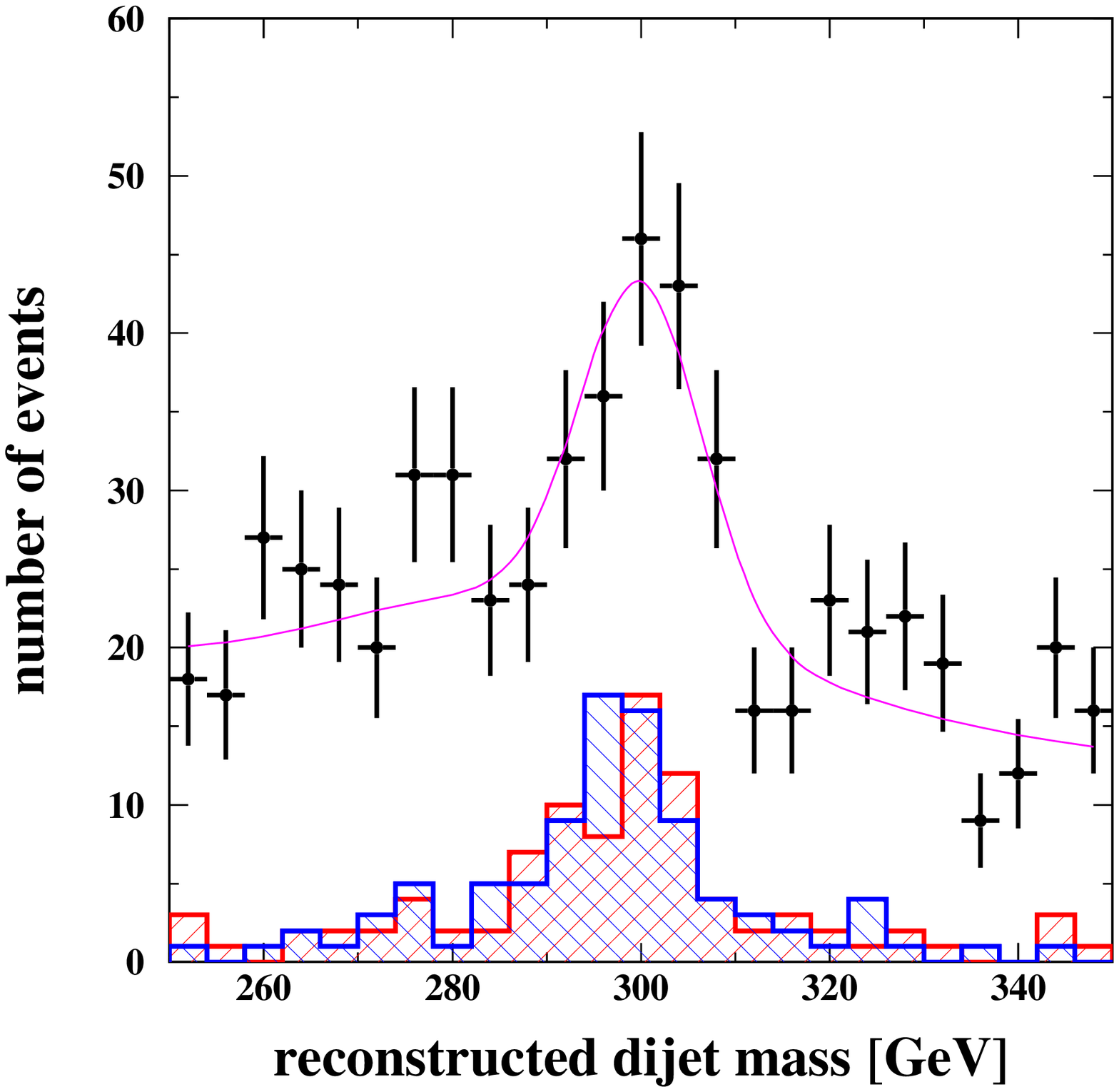}\\
\vspace{-10mm}
\includegraphics*[width=0.7\textwidth]{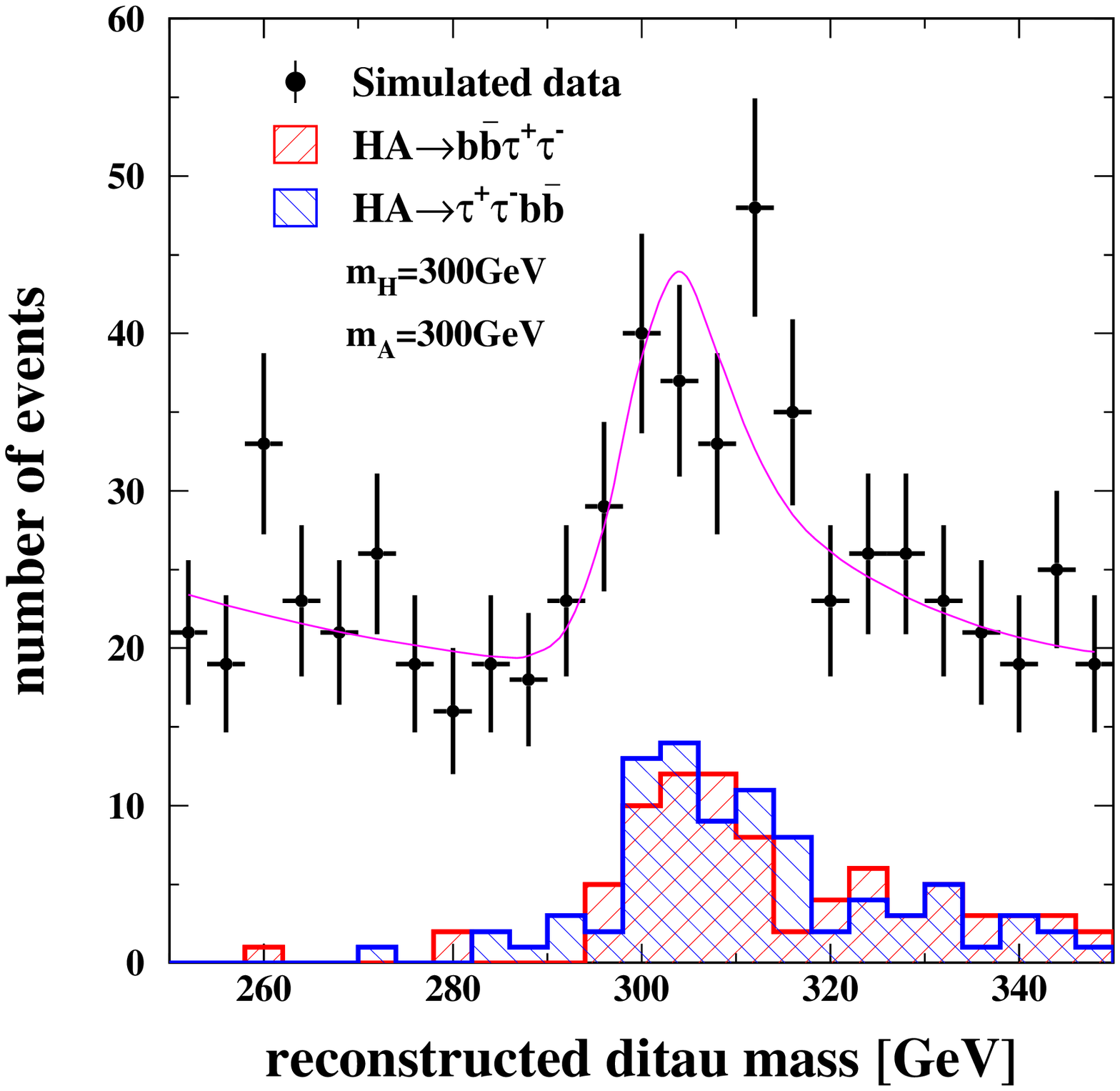}
\end{center}
\caption{
Distributions of di-jet mass and di-tau mass in the $\HAbbttbb$ channels for the case of ($\mH$,$\mA$) = (300,300) GeV at $\sqrts$ = 800 GeV.
\label{fig:mass_bbtt_eq}
}
\end{figure}

In the $\HAbbttbb$ channels, in the case of large mass splitting between 
Higgs bosons, the signal exhibits itself as two peaks in the 
reconstructed di-jet and di-tau mass spectra as illustrated in 
Fig.~\ref{fig:mass_bbtt}.
The spectra are fitted simultaneously with the superposition of background
and signal distributions. The fit is performed with 
four free parameters : two normalization factors for the 
$\HAbbtt$ and $\HAttbb$ samples and two Higgs boson masses. 
Background distribution is assumed to be well measured and therefore fixed.
The shapes of signal peaks are parametrised using high statistics
MC samples. When H and A are degenerate in mass, 
signal distributions overlap and 
the contributions from the $\HAbbtt$ and $\HAttbb$ samples 
cannot be disentangled. To stabilize the fitting procedure we assume that
the numbers of the $\HAbbtt$ and $\HAttbb$ events, selected 
into the final sample, are equal.
This assumption is validated by the properties of the 
decoupling limit. As was mentioned in the introductory section,
in the decoupling limit the H and A bosons have almost identical
properties: $\mH$ $\approx$ $\mA$, $\BrHbb$ $\approx$ $\BrAbb$, 
$\BrHtt$ $\approx$ $\BrAtt$, $\gH$ $\approx$ $\gA$. 
Consequently, the $\HAbbtt$ and $\HAttbb$ events are expected  
to contribute equally to the final selected sample. This
reduces the number of free parameters to three : a common normalization factor for the two signal samples and two Higgs
boson masses. As an example, Fig.~\ref{fig:mass_bbtt_eq} 
shows the result of the fit for 
($\mH$,$\mA$) = (300,300)~\GeV. 

Table~\ref{tab:prec_mass} summarizes the 
statistical accuracy of the mass measurements for the $\HAbbbb$ and $\HAbbttbb$ channels and their combination. The statistical accuracy of the combined 
mass measurement is $\sim$ 100 MeV for the Higgs pair production far
above kinematic threshold and degrades to $\sim$ 1 GeV with aproaching
kinematic limit. At the same time the statistical errors for the $\sqrts$ = 800 GeV are in general twice as bigger than those for the 500 GeV for the both channels. The combination of the two channels helps to improve the mass determination accuracy.    

The statistical uncertainty on the topological cross 
sections is reported in Table~\ref{tab:prec_xsec}. In the $\HAbbbb$ channel, the topological 
cross sections can be measured with relative precision between 1.5 and 6.6\% while in the \newline $\HAbbttbb$ channels between 5 to 30\%. 

\begin{figure}
\begin{center}
\includegraphics*[width=0.7\textwidth]{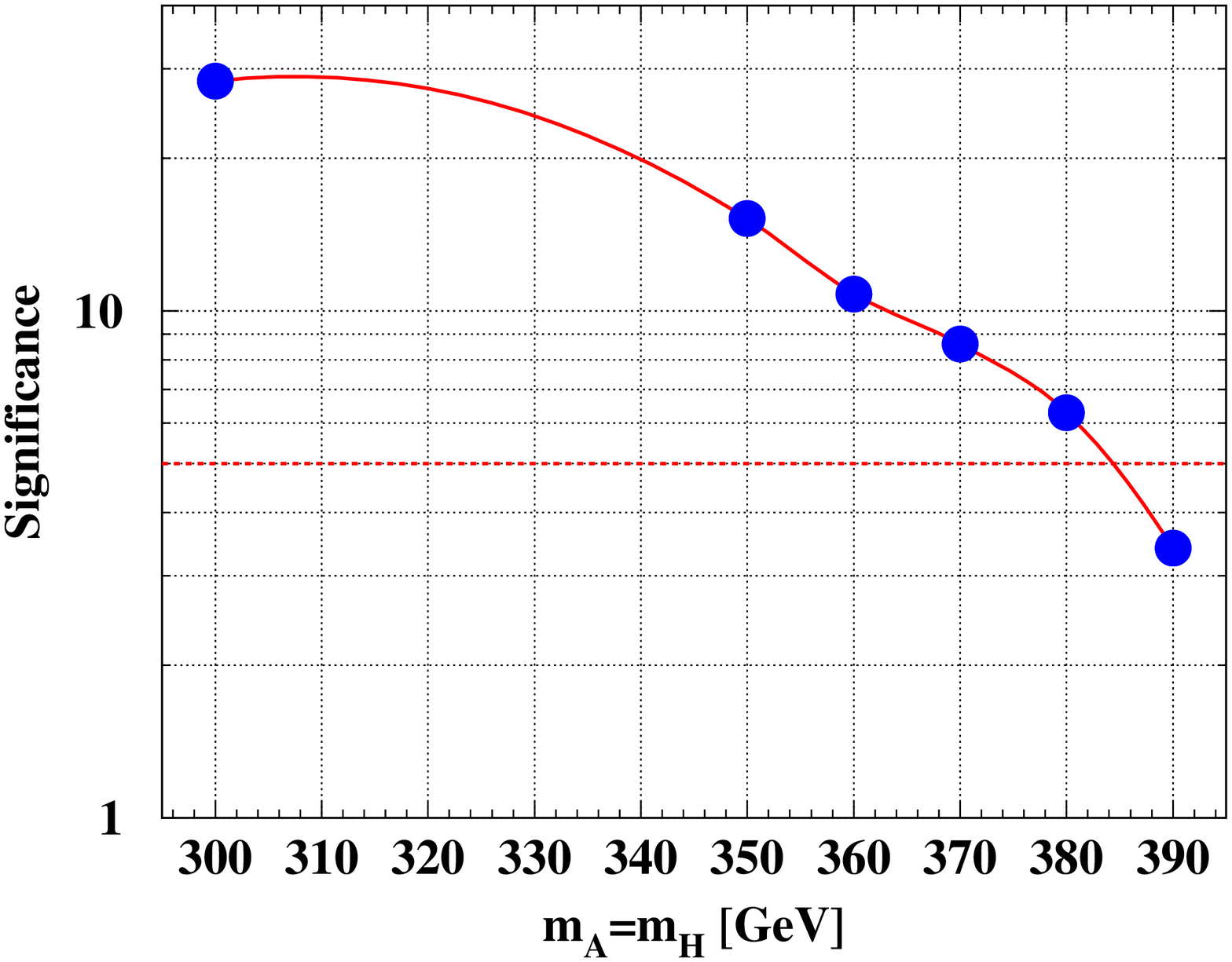}
\hspace{-8mm}
\includegraphics*[width=0.7\textwidth]{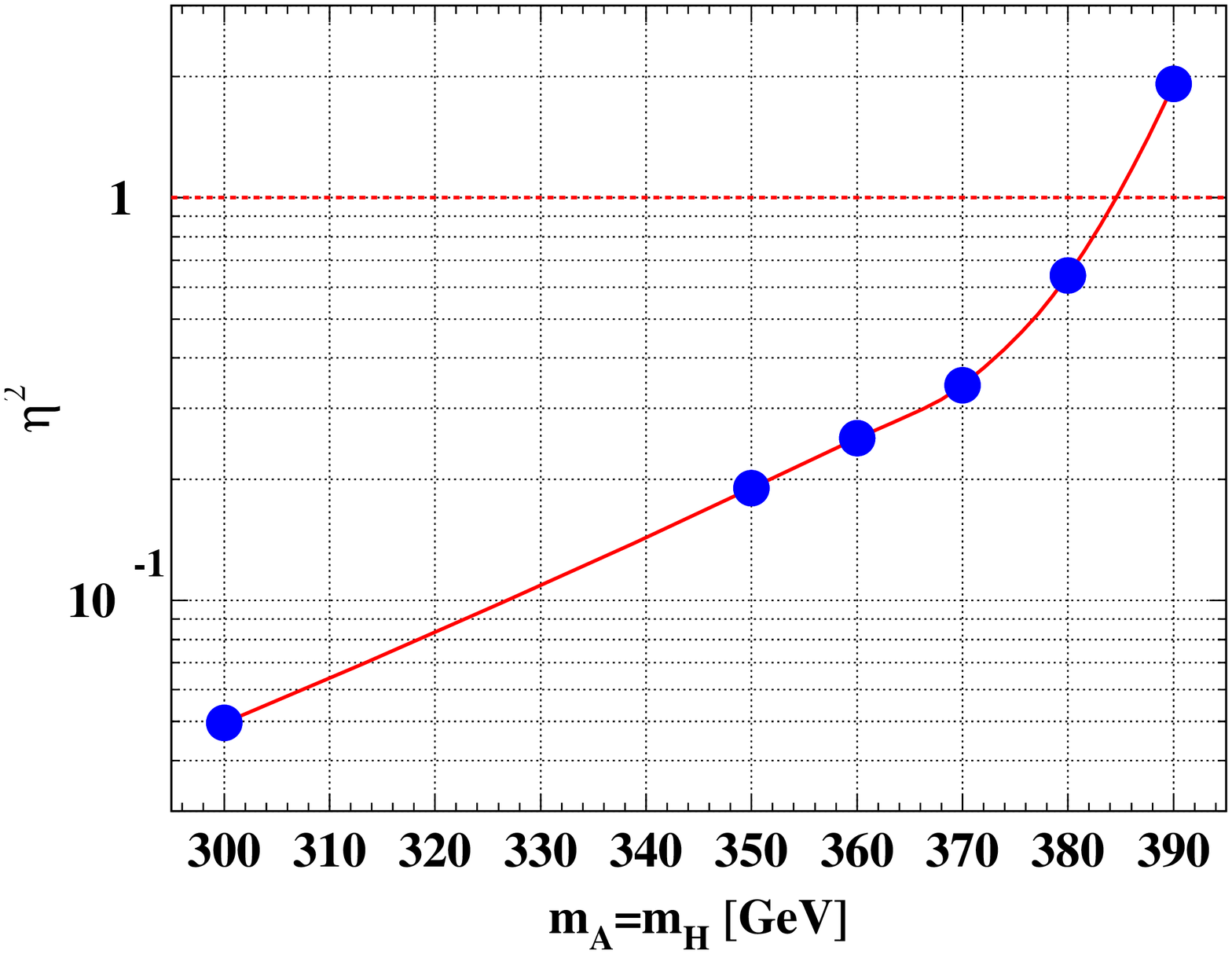}
\end{center}
\caption{
Discovery significance as a function of the Higgs boson mass (assuming $m_{H}=m_{A}$) in the $\HAbbbb$ channel at $\sqrts$ = 800 GeV (upper figure). $\eta^2$ as a function of the Higgs boson mass for the 5$\sigma$ discovery limit (lower figure) where $\eta^2$ is assumed $\eeHA$ cross section relative to that for $\sin^2(\beta-\alpha) = 1$. Branching ratios 
$\BrHbb$ and $\BrAbb$ are assumed to be 90\%.
\label{fig:signific}
}
\end{figure}

A big part of SUSY parameter space leads to degenerate H and A Higgs boson masses. For this case the discovery significance as a function of $m_H$ (=$m_A$) (Fig.~\ref{fig:signific}) is calculated for the $\HAbbbb$ channel at $\sqrts$ = 800 GeV. Approaching the kinematic limit, the significance drops below 5$\sigma$ between 380 GeV to 390. The whole range of the significances for the Higgs masses is from 28.2 to 3.4. The 5$\sigma$ discovery reach and $\eta^2$ as a function of $m_H$ (=$m_A$) where $\eta^2$ is assumed $\eeHA$ cross section relative to that for $\sin^2(\beta-\alpha) = 1$ (as shown in Fig.~\ref{fig:signific}).

\subsection{Width}

\begin{figure}
\begin{center}
\includegraphics*[width=0.7\textwidth]{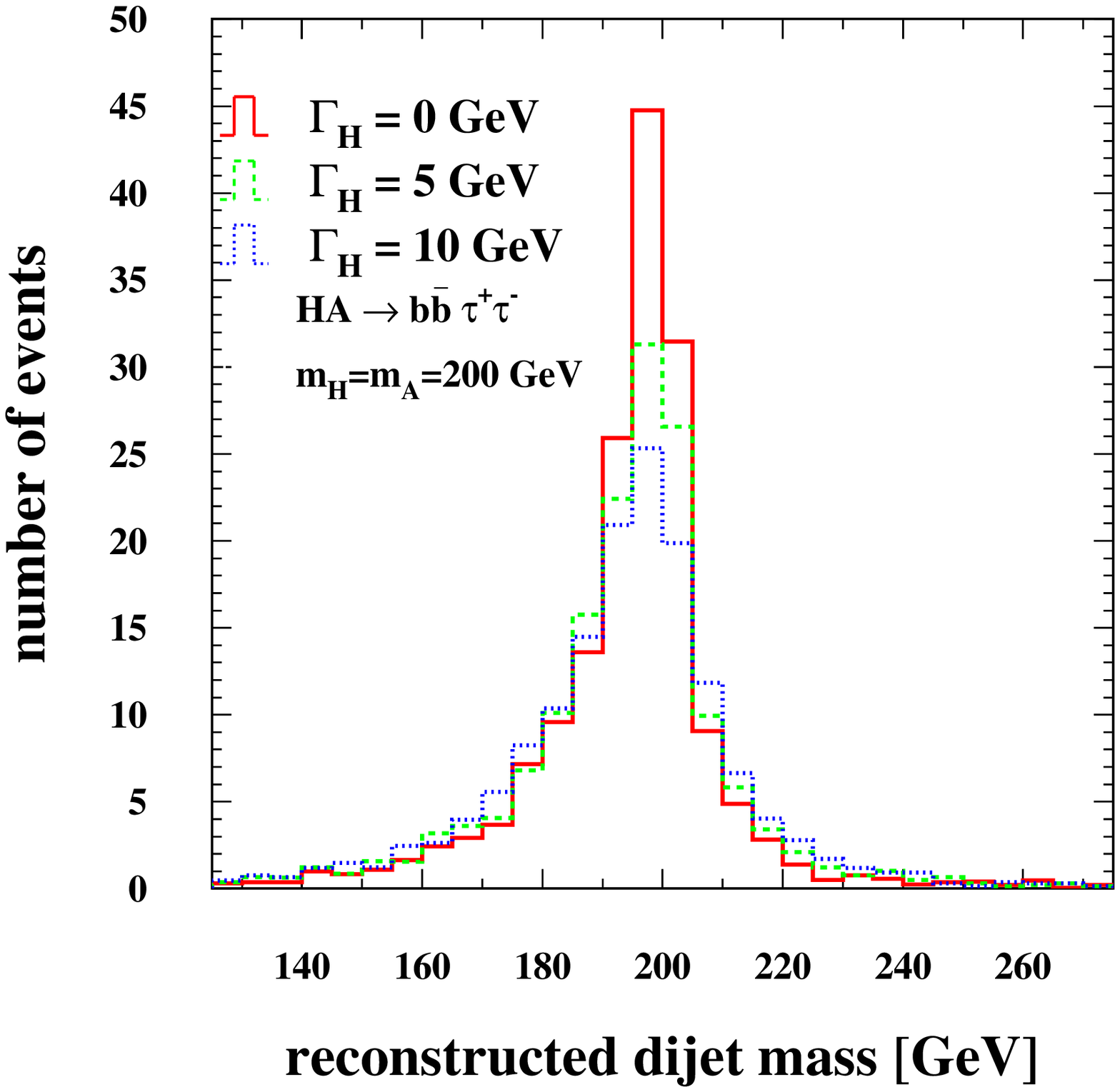}\\
\vspace{-10mm}
\includegraphics*[width=0.7\textwidth]{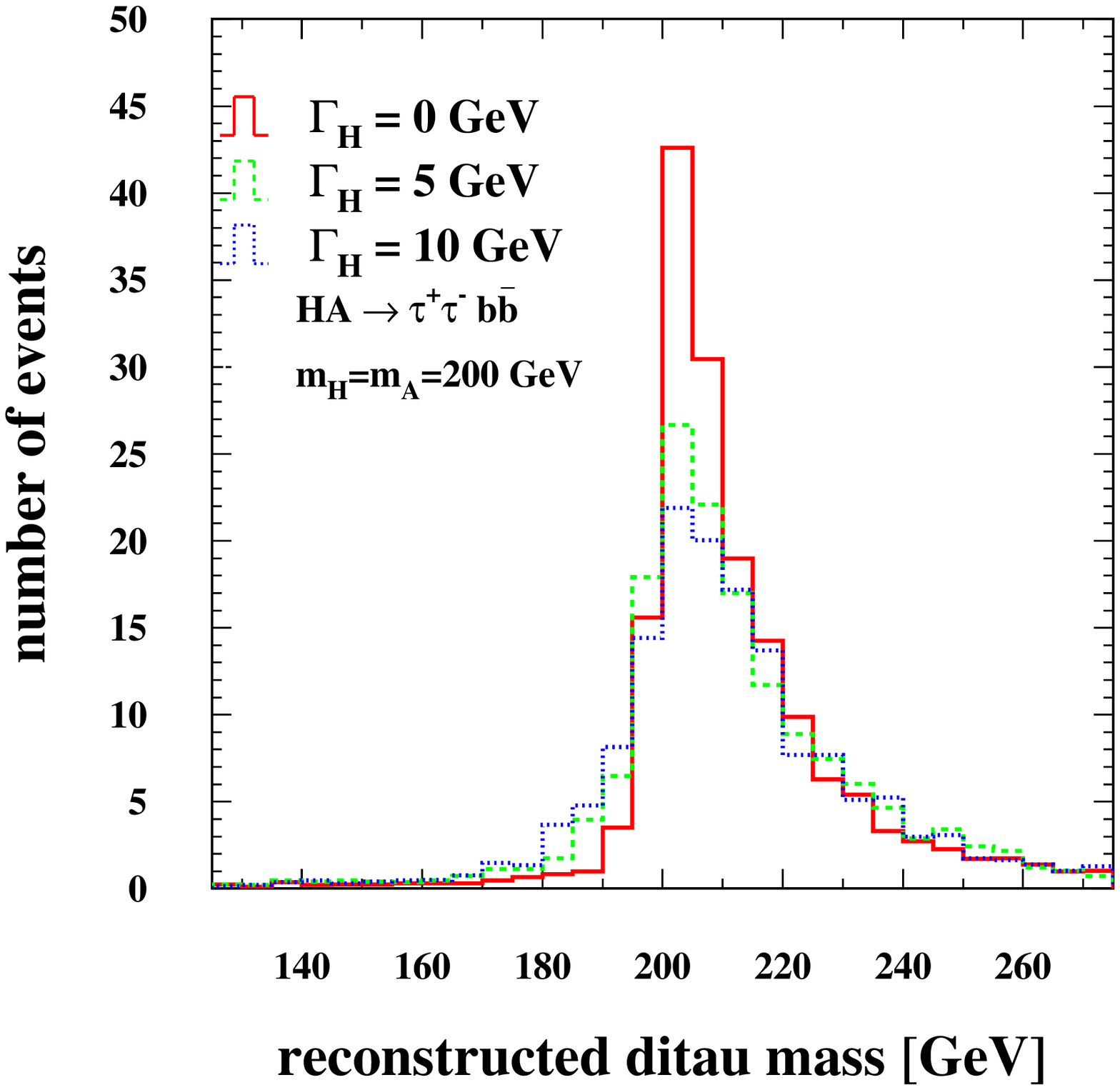}
\end{center}
\vspace{-4mm}
\caption{
The reconstructed di-jet mass spectrum in the $\HAbbtt$ sample (upper figure) 
and the reconstructed di-tau mass spectrum in the $\HAttbb$ sample (lower figure)
for the different Higgs boson widths $\gH$ = 0, 5, 10 GeV for $m_{H,A}$ = 200 GeV at $\sqrts$ = 500 GeV.
\label{fig:width_impact}
}
\end{figure}

\begin{figure}
\begin{center}
\includegraphics*[width=0.7\textwidth]{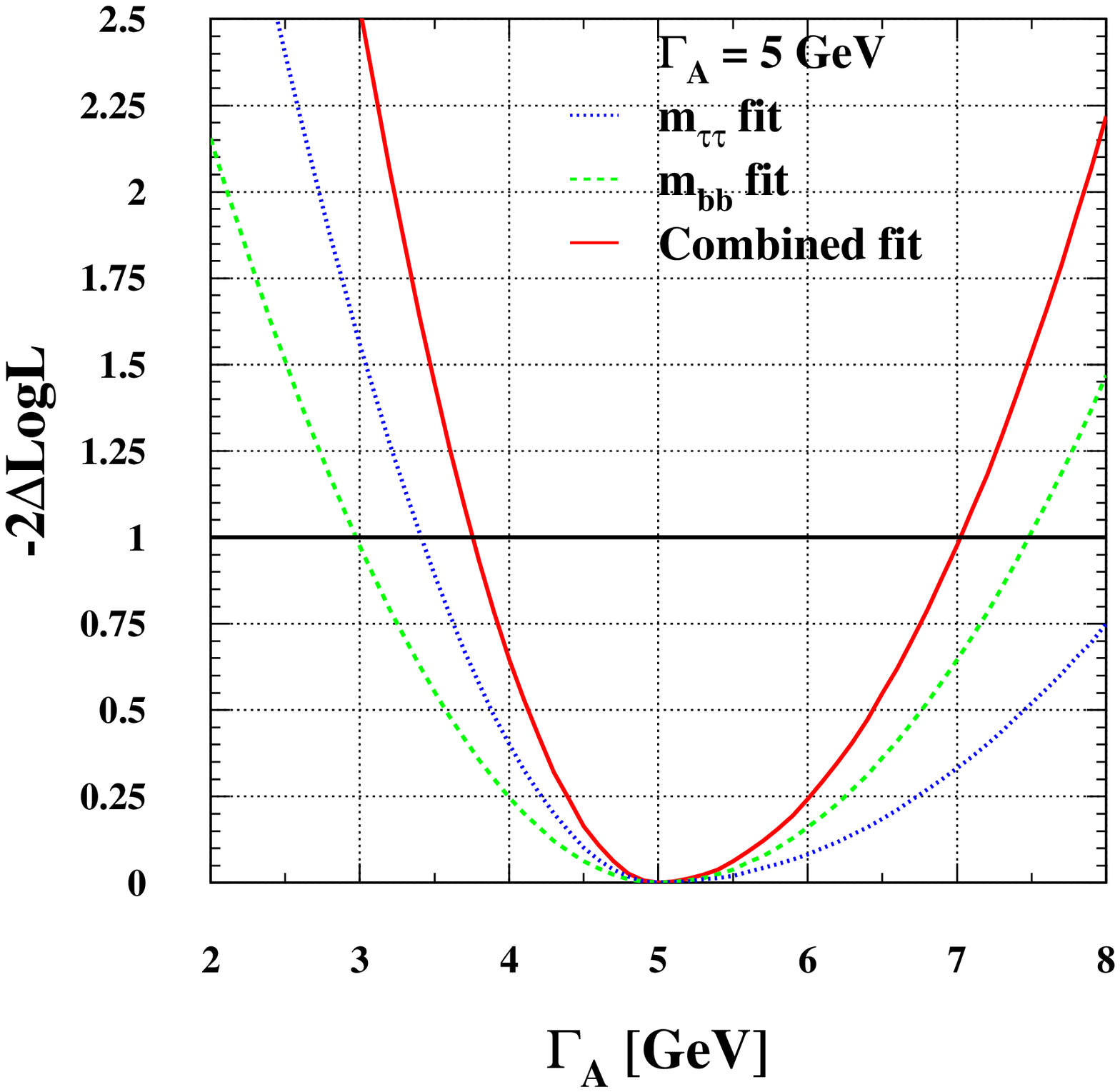}\\
\vspace{-10mm}
\includegraphics*[width=0.7\textwidth]{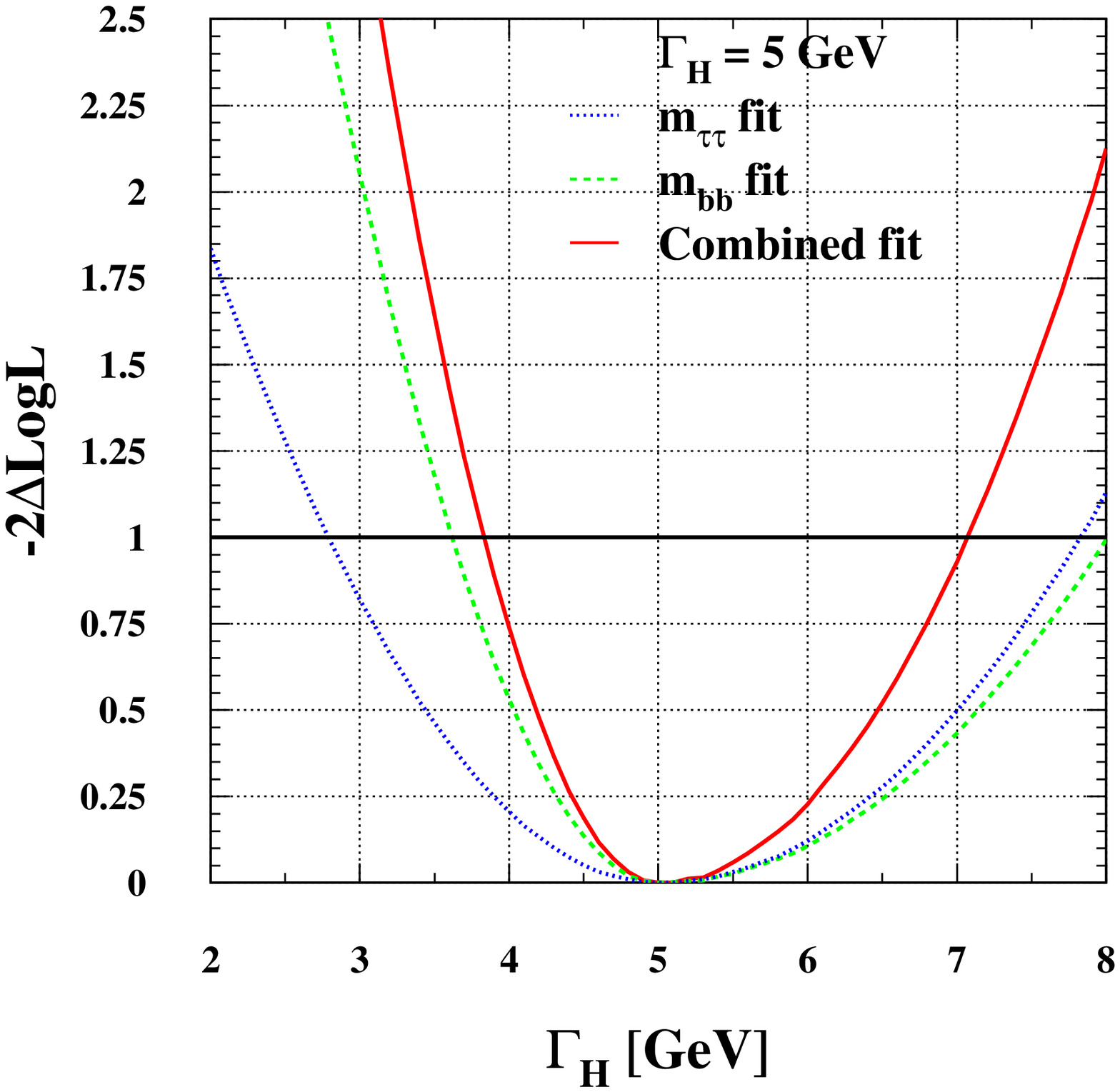}
\end{center}
\caption{
The log-likelihood as a function of $\gH$ and $\gA$ for the case
of $\gH = \gA$ = 5 GeV and ($\mH$,$\mA$) = (200,150) GeV at $\sqrts$ = 500 GeV.
\label{fig:width_llr}
}
\end{figure}

We utilize the $\HAbbttbb$ channels to measure the Higgs boson widths.
Fig.~\ref{fig:width_impact}  illustrates the impact of the natural widths of 
the Higgs bosons on the reconstructed di-jet and di-tau mass spectra. 
We consider separately two cases :
\begin{enumerate}
\item large mass splitting between H and A,
\item mass degeneracy. 
\end{enumerate}
In the first case, the strategy of the measurement is the following. 
The masses of Higgs bosons are fixed to the values measured in the 
$\HAbbbb$ channel.
The reconstructed di-jet and di-tau mass line shapes are parametrised 
as a function 
of $\gH$ and $\gA$ using MC samples of the $\HAbbtt$ and 
$\HAttbb$ events generated with different Higgs boson widths. A simultaneous 
log-likelihood fit of the parametrised mass line shapes 
to the actual reconstructed di-jet and di-tau mass distributions is performed
with four free parameters : $\gH$, $\gA$
and two normalization factors for the $\HAbbtt$ and $\HAttbb$ samples.
Fig.~\ref{fig:width_llr} shows the dependence of the log-likelihood 
on the probed Higgs boson widths for the case of $\gH$ = $\gA$ = 5 GeV and
($\mH$,$\mA$) = (200,150) GeV at $\sqrts$ = 500 GeV. The dependence is shown for the case when only di-jet mass or di-tau mass spectrum are fitted and for the simultaneous fit to the two distributions. The feasible accuracies of the 
width measurements for three representative Higgs boson mass hypotheses 
are given in Table~\ref{tab:width1}. The precision on the width determination ranges from 600 MeV to 4.0 GeV.

In the case of very close Higgs boson masses, the broadening 
of the signal peak in the mass distributions can be caused by two factors :
the finite natural widths of the Higgs bosons and non-zero mass 
difference, $\delta M = \mH-\mA$. The width measurement in this case proceeds
as follows. The sum of the Higgs boson masses is fixed to the value 
measured in the $\HAbbbb$ channel, $\mH+\mA=2M$. 
From the $\HAbbtt$ and $\HAttbb$ MC 
samples with $\gH$ = $\gA$ = 0 GeV, the experimental mass resolution functions 
$R_{bb,\tau\tau}(x)$ are derived, which approximate the reconstructed 
di-jet and di-tau mass distributions in the limit of infinitely small 
Higgs widths. To ensure stability of the fitting procedure, the H and A
boson are assumed to have identical decay properties,
$\BrHbb$ = $\BrAbb$, $\BrHtt$ = $\BrAtt$, $\gH$ = $\gA$ = $\Gamma$, 
as expected in the decoupling limit.
The distributions of the invariant di-jet and di-tau 
masses are fitted simultaneously with the functions
\begin{displaymath}
\begin{array}{ccccccc}
F_{bb,\tau\tau}(x) & = & 
B_{bb,\tau\tau}(x) & + & 
N\cdot BW(x,M\pm\delta M/2,\Gamma)\otimes R_{bb,\tau\tau}(x) & + \\
& & & & 
N\cdot BW(x,M\mp\delta M/2,\Gamma)\otimes R_{bb,\tau\tau}(x) & \\
\end{array}
\end{displaymath}
The fitting functions consist of three terms. The first one, 
$B_{bb,\tau\tau}(x)$, describes the background spectrum. The other 
two describe two signal distributions and represent 
the convolution of the Breit-Wigner function 
\begin{displaymath}
BW(x,M,\Gamma) = \frac{xM\Gamma}{(x^2-M^2)^2+M^2\Gamma^2}
\end{displaymath}
with the experimental resolution functions $R_{bb,\tau\tau}(x)$. 
The fit is performed with the common normalization factor, $N$,
the Higgs boson width, $\Gamma$, and the Higgs boson mass difference, 
$\delta M$, as free parameters. The Higgs width errors
obtained from the fitting procedure are presented in 
Table~\ref{tab:width2} for three representative Higgs mass hypotheses. The precision on the width determination ranges from 1 GeV to 4.5 GeV.

\section{Conclusion}
We examined the potential of a future linear collider detector for the 
determination of the MSSM Higgs boson properties exploiting 
associated Higgs boson pair production followed by the Higgs decays
to b-quarks and $\tau$-leptons. It is shown that combining 
the $\HAbbbb$ and $\HAbbttbb$ channels the Higgs boson 
masses can be measured with an accuracy of up to several hundred MeV for 
Higgs pair production far above the kinematic threshold. 
The precision deteriorates to about 1 GeV with approaching the kinematic
threshold. The topological
cross section $\sHAbbbb$ can be measured with a relative precision 
varying between 1.5 and 6.6\%. The error on the topological cross sections 
$\sHAbbtt$ and \newline $\sHAttbb$ is estimated to range between 4 and 30\%. 
Moderate constraints on the Higgs boson widths can be set from 
the analysis of the reconstructed di-jet and di-tau mass spectra in the
$\HAbbtt$ and $\HAttbb$ channels.

The 5$\sigma$ discovery limit corresponds to the Higgs mass of around 
385 GeV for the degenerate Higgs boson masses in 
the $\HAbbbb$ channel at $\sqrts$ = 800 GeV with integrated luminosity
of 500 fb$^{-1}$. 

\newpage 


\appendix
\section{SPS 1a}
The present analysis is applied to one of the so-called benchmark points SPS 1a for SUSY searches~\cite{SPS}. SPS 1 is a typical mSUGRA scenario which consists of a point with an intermediate value of $\tan \beta$ and a model line attached to it (SPS 1a) and of a "typical" mSUGRA point with relatively high $\tan \beta$ (SPS 1b). The parameters for the SPS 1a point are $m_0$ = 100 GeV, $m_{1/2}$ = 250 GeV, $A_{0}$ = -100 GeV,  $\tan \beta$ = 10,  $\mu >0$. For this point the Higgs masses are $m_{h^0}$ = 113.7 GeV, $m_{A^0}$ = 394.65 GeV, $m_{H^0}$ = 394.9, $m_{H^{\pm}}$ = 403.6 GeV according to the Hdecay and Feynhiggsfast programs~\cite{Hdecay,Feynhiggsfast}. 

The analysis is done for the center-of-mass energy of $\sqrt{s}=1\,\mathrm{TeV}$, at which the cross section for the process $\eeHA$ is 2.5 fb. The luminosity assumed is 1000 fb$^{-1}$. The branching ratio for the H (A) Higgs boson to $b\bar{b}$ is 0.64 (0.40), $\Gamma_{tot}$ = 0.785 GeV (1.251 GeV). 

\begin{figure}
\begin{center}
\includegraphics*[width=0.7\textwidth]{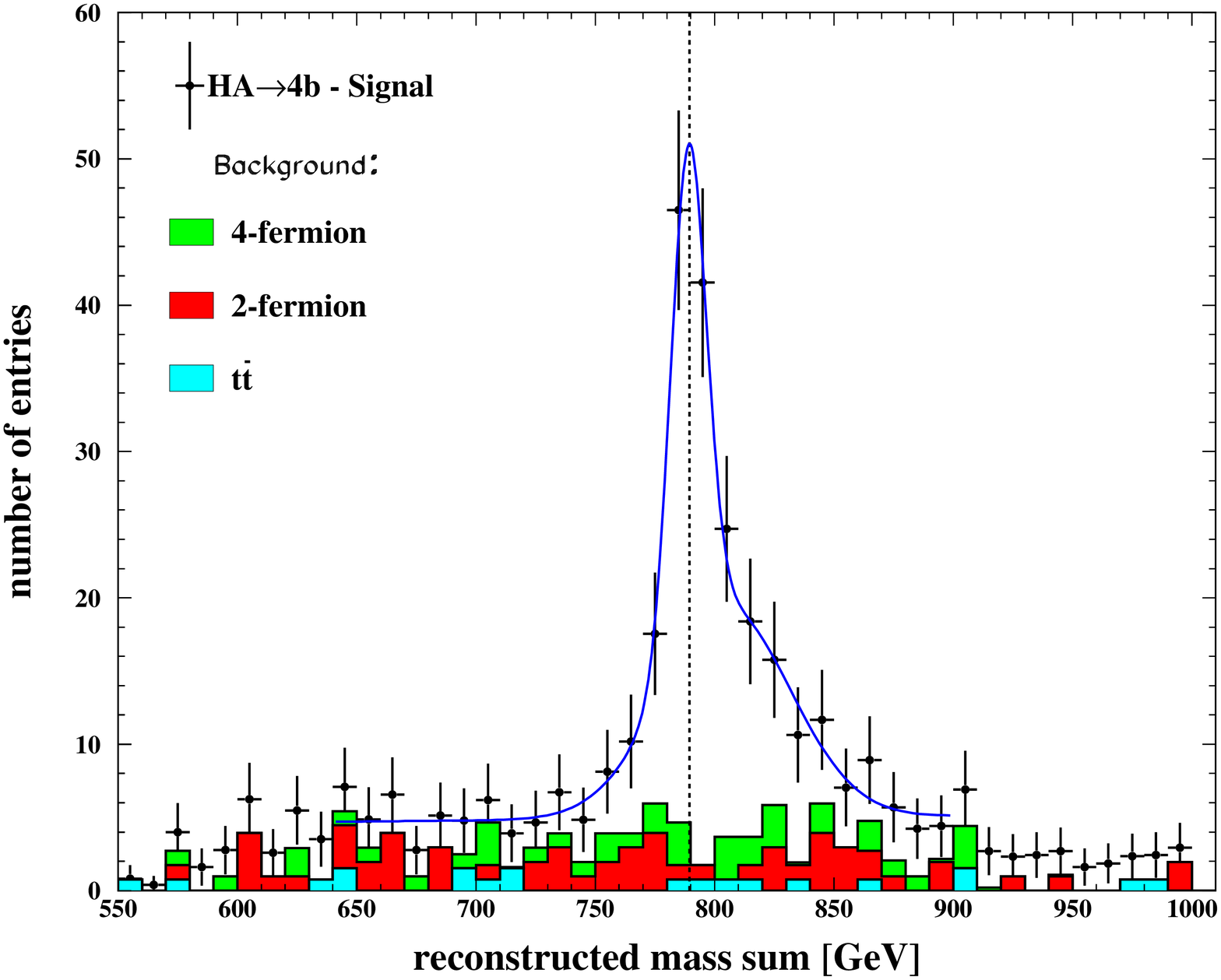}
\hspace{-8mm}
\includegraphics*[width=0.7\textwidth]{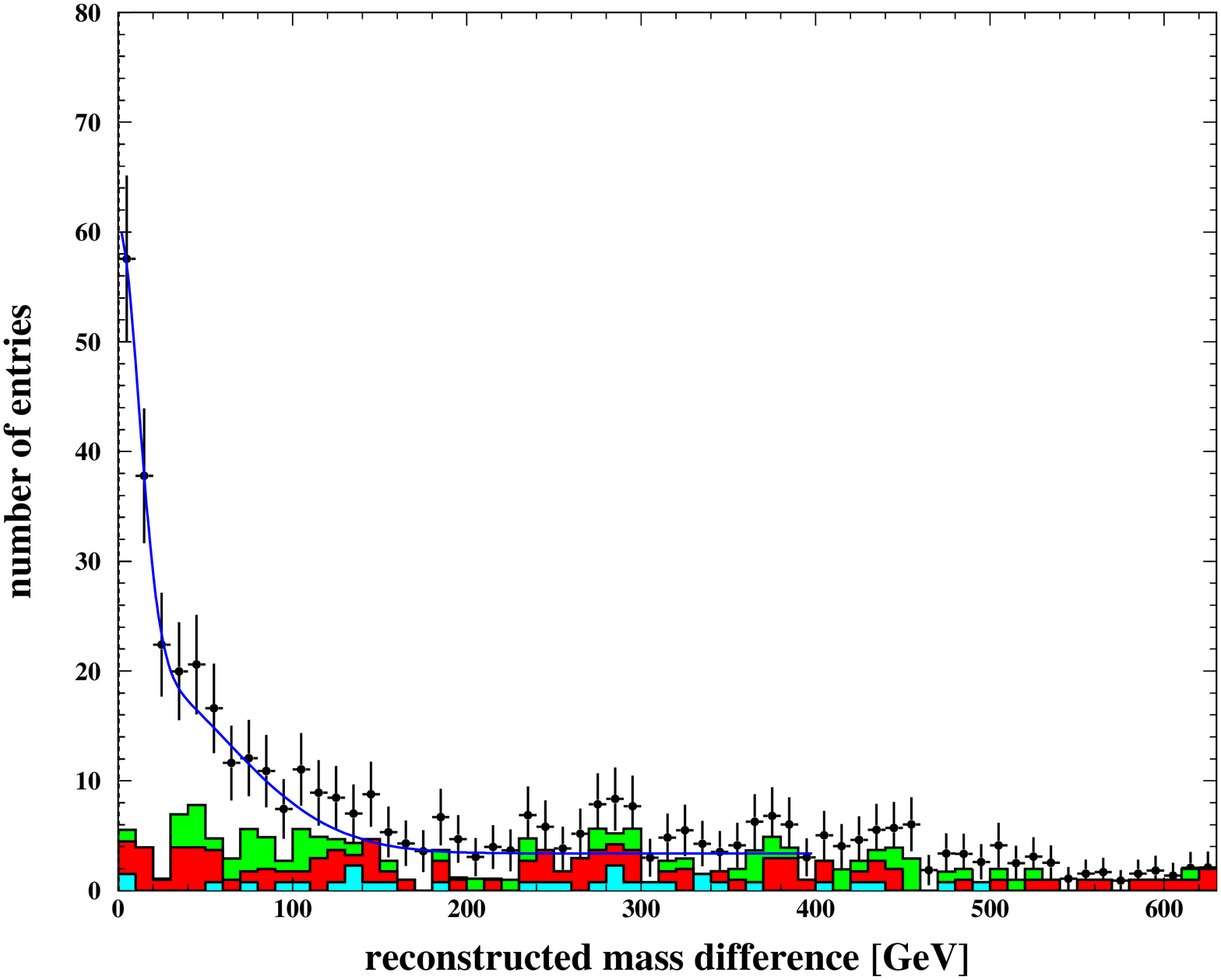}
\end{center}
\caption{
Upper figure: distribution of the di-jet mass sum after selection cuts, kinematic fit and cut on di-jet mass difference. Lower figure: distribution of the di-jet mass difference
after selection cuts, kinematic fit and cut on di-jet mass sum. Both distributions are in the $\HAbbbb$ channel for the SPS 1a benchmark point with Higgs boson mass hypothesis ($\mH$,$\mA$) = (394.90,394.65) GeV at $\sqrts$ = 1 TeV.
\label{fig:sps}
}
\end{figure}

The results presented in Fig.~\ref{fig:sps} are the mass sum and the mass difference after the selection cuts, kinematic fit and the final cuts for the mass difference and the mass sum respectively. The masses can be measured with precision of 1.3 GeV. The signal efficiency is 29\% after selection cuts and 24\% after cuts on di-jet mass sum and difference. The cross section can be measured with the relative uncertainty of 9\%.   

\newpage 

\bibliographystyle{unsrt}

\newpage


\begin{table}[bh]
\begin{center}
\begin{tabular}{|c|c|c|}
\hline
$\sqrts$ [GeV]& ($\mH$,$\mA$) [\GeV] & $\sHA$ [fb] \\
\hline \hline
   &(150,100)      &     33.62        \\
   &(200,100)      &     25.30        \\
   &(250,100)      &     16.61        \\
500   &(150,140)      &     28.39        \\
&(150,150)      &     26.90        \\
   &(200,150)      &     18.85        \\
   &(250,150)      &     10.67        \\
   &(200,200)      &     11.35        \\
\hline
   &(300,150)      &     10.55  \\
   &(290,200)      &      9.54  \\
   &(300,250)      &      7.49  \\
800&(300,300)      &      5.70  \\
   &(350,350)      &      2.23  \\
   &(400,150)      &      6.46  \\
   &(400,200)      &      5.17  \\
   &(400,250)      &      3.70  \\
\hline
\end{tabular}
\end{center}
\caption{
Tree level cross sections $\sHA$ for $\eeHA$ expected for the Higgs boson mass 
hypotheses ($\mH$,$\mA$) considered in the study.
Numbers are given for $\sin^2(\beta-\alpha)$=1. Cross sections 
are calculated including ISR at center-of-mass energies of 
$\sqrts$ = 500, 800 GeV. 
\label{tab:higgs_sample}
}
\end{table}

\begin{table}
\begin{center}
\begin{tabular}{|c|c|c|}
\hline
Process  & \multicolumn{2}{|c|}{$\sigma \times BR$ [fb]} \\ \cline{2-3}
         & 500 \GeV & 800 \GeV \\
\hline \hline
$\ttWbWb$  & $6.69\cdot 10^2$ & $1.65\cdot 10^2$ \\
$\WWqqqq$  & $4.13\cdot 10^3$ & $2.34\cdot 10^3$ \\
$\WWqqln$  & $3.88\cdot 10^3$ & $2.20\cdot 10^3$ \\
$\ZZqqqq$  & $3.11\cdot 10^2$ & $1.74\cdot 10^2$ \\
$\ZZqqll$  & $0.89\cdot 10^2$ & $4.97\cdot 10^1$ \\
$\ZZqqnn$  & $1.78\cdot 10^2$ & $9.96\cdot 10^1$ \\
$\qqal$    & $1.29\cdot 10^4$ & $5.45\cdot 10^3$ \\
$\Wenqq$   & $5.08\cdot 10^3$ & $8.40\cdot 10^3$ \\
$\Zeeqq$   & $4.46\cdot 10^3$ & $4.10\cdot 10^3$ \\
\hline
\end{tabular}
\end{center}
\caption{
Topological cross sections $\sigma \times BR$ of background processes
at \newline $\sqrts$ = 500, 800~\GeV.
\label{tab:backgrounds}
}
\end{table}

\begin{table}[p]
\begin{center}
\begin{tabular}{{|c||c||c|c|c|c||c|c|}} \hline 
Cut& Tot. BG &   ZZ   &   WW    &  2-ferm.&$t\bar{t}$&Eff.\% & Signal \\ \hline \hline
   & 9380780 & 157240 & 2087383 & 6799865 & 336290   & 100.0 & 13618  \\  
1  & 6035870 & 146909 & 1957360 & 3651766 & 279839   & 98.27 & 13383  \\
2  & 2675300 & 105981 & 1309760 & 1047851 & 211705   & 91.34 & 12438  \\
3  & 1357080 & 39789  & 383774  & 772600  & 160915   & 85.06 & 11583  \\
4  & 335995  & 8167   & 50257   & 119448  & 158123   & 67.81 & 9235   \\
5  & 306623  & 7832   & 48523   & 92288   & 157981   & 66.58 & 9067   \\
6  & 265119  & 7499   & 46254   & 84886   & 126480   & 63.18 & 8604   \\
7  & 148085  & 5975   & 35270   & 70218   & 36621    & 54.39 & 7407   \\
8  & 41912   & 1761   & 809     & 10136   & 29206    & 53.22 & 7247   \\ \hline
9  & 1960    & 272    & 19      & 473     & 1196     & 42.63 & 5805   \\ 
\hline
\end{tabular}
\end{center}
\caption{Number of remaining background events, signal efficiency and number of expected signal 
events after each of the nine cuts for the mass point ($\mH$,$\mA$) = (150,100) in the $\HAbbbb$ channel at $\sqrts$ = 500 GeV. The first column is the number of the cut, the second one is the number of total background events. 
\label{tab:cutflow_bbbb_500}
}
\end{table}

\begin{table}[p]
\begin{center}
\begin{tabular}{{|c||c||c|c|c|c||c|c|}} \hline 
Cut& Tot. BG &   ZZ   &   WW    &  2-ferm.&$t\bar{t}$&Eff.\% & Signal \\ \hline \hline
   & 4137520 & 87211 & 1183901& 2783826 & 82579   &100.0& 3040  \\  
1  & 2376220 & 69491 & 922289 & 1307850 & 76592   &97.00& 2949  \\
2  & 1007200 & 41522 & 462295 & 432605  & 70781   &93.98& 2857  \\
3  & 542191  & 15305 & 154934 & 319818  & 52133   &85.67& 2604  \\
4  & 90601   & 2328  & 21986  & 42023   & 24265   &82.22& 2499   \\
5  & 78613   & 2117  & 20732  & 31502   & 24262   &80.29& 2441   \\
6  & 54801   & 1937  & 16460  & 29276   & 7128    &71.31& 2168   \\
7  & 12283   & 534   & 2075   & 4430    & 5244    &68.99& 2097 \\ \hline
8  & 488     & 50    & 57     & 175     & 206&44.92&   1366 \\ \hline
\end{tabular}
\end{center}
\caption{Number of remaining background events, signal efficiency and number of expected signal events after each of the eight cuts for the mass point ($\mH$,$\mA$) = (300,250) in the $\HAbbbb$ channel at $\sqrts$ = 800 GeV. The first column is the number of the cut, the second one is the number of total background events. 
\label{tab:cutflow_bbbb_800}
}
\end{table}

\begin{table}
\begin{center}
\begin{tabular}{|c|c|c|c|c|c|c|c|}
\hline
$\sqrts$ [GeV]&($\mH$,$\mA$) [GeV] &\multicolumn{2}{|c|}{Efficiency [\%]}&\multicolumn{2}{|c|}{ Number of events}& \multicolumn{2}{|c|}{Tot. backgr.}\\ \cline{3-8}
 &  & cuts& 2d.cut& cuts& 2d.cut&cuts& 2d.cut\\
\hline \hline
   &(150,100)      &  43 &31&  5805& 4196 &    &132 \\
   &(200,100)      &  41 &26&  4194& 2661 &    &129  \\
   &(250,100)      &  36 &28&  2422& 1851 &    &182  \\
500   &(150,140)      &  48 &39&  5469& 4518 &1960    &301  \\
   &(150,150)      &  45 &41&  4944& 4407 & &144    \\
   &(200,150)      &  42 &27&  3192& 2036 &    &130 \\
   &(250,150)      &  36 &22&  1534& 960  &    &156 \\
   &(200,200)      &  37 &33&  1691& 1510 &    &185 \\
\hline
   &(300,150)      & 43  &33& 1825 & 1427&    & 38\\
   &(290,200)      & 46  &36& 1763 & 1398&    & 53\\
   &(300,250)      & 45  &36& 1366 & 1101&    & 13\\
800&(300,300)      & 44  &35& 1011 & 823 & 488& 26\\
   &(350,350)      & 37  &31&  339 & 277 &    & 58\\
   &(400,150)      & 37  &26&  969 & 678 &    & 27\\
   &(400,200)      & 40  &28&  833 & 584 &    & 41\\
   &(400,250)      & 40  &28&  597 & 414 &    & 49\\
\hline
\end{tabular}
\end{center}
\caption{
The signal efficiencies, the number of signal and total background events after selection cuts and after cuts on di-jet mass sum and difference in the $\HAbbbb$ channel at $\sqrts$ = 500 and 800~\GeV. 
The signal expectations 
are quoted for $\sin^2(\beta-\alpha)$ = 1 and the Higgs boson 
branching fractions of $\BrHbb$ = $\BrAbb$ = 90\%.
\label{tab:bbbb_effy}
}
\end{table}

\begin{table}
\begin{center}
\begin{tabular}{|c|c|c|c|c|c|c|}
\hline
$\sqrts$ [GeV]&($\mH$,$\mA$) [GeV] & \multicolumn{2}{c|}{$\HAbbtt$} & \multicolumn{2}{c|}{$\HAttbb$} & Tot.BG \\  \cline{3-6}
  &                 & Eff.[\%]& Events&Eff.[\%]& Events &\\
\hline \hline
      &(150,100)      &  68 & 1035 & 73 & 1110 &         \\
      &(200,100)      &  67 &  765 & 72 &  820 &         \\
      &(250,100)      &  61 &  460 & 70 &  525 &         \\
500   &(150,140)      &  73 &  940 & 73 &  940 & 133490        \\
      &(150,150)      &  74 &  900 & 74 &  900 &         \\
      &(200,150)      &  70 &  595 & 71 &  605 &         \\
      &(250,150)      &  63 &  305 & 70 &  340 &         \\
      &(200,200)      &  69 &  355 & 69 &  355 &         \\
\hline
   &(300,150)  & 60 & 290 & 72 & 345 &\\
   &(290,200)  & 65 & 280 & 70 & 305 &\\
   &(300,250)  & 64 & 215 & 67 & 230 &\\
800&(300,300)  & 64 & 165 & 64 & 165 &52020\\
   &(350,350)  & 58 &  55 & 58 &  55 &\\
   &(400,150)  & 52 & 150 & 71 & 205 & \\
   &(400,200)  & 55 & 130 & 68 & 160 &\\
   &(400,250)  & 56 &  95 & 66 & 110 &\\
\hline
\end{tabular}
\end{center}
\caption{
The $\HAbbtt$ and $\HAttbb$ efficiencies, the number of signal and total background events after preselection at $\sqrts$ = 500 and 800~\GeV. The signal expectations 
are quoted for $\sin^2(\beta-\alpha)$ = 1 and the Higgs boson 
branching fractions of $\BrHbb$ = $\BrAbb$ = 90\%, 
$\BrHtt$ = $\BrAtt$ = 10\%.
\label{tab:sgnl_pres_bbtt}
}
\end{table}

\begin{table}
\begin{center}
\begin{tabular}{|c|c|c|c|c|c|c|}
\hline
$\sqrts$ [GeV]&($\mH$,$\mA$) [GeV] & \multicolumn{2}{c|}{$\HAbbtt$} & \multicolumn{2}{c|}{$\HAttbb$}& Tot.BG  \\  \cline{3-6}
  &                 & Eff.[\%]& Events & Eff.[\%]& Events& \\
\hline \hline
   &(150,100)      &   42  &  640 & 44  &  670 & 5070  \\
   &(200,100)      &   39  &  445 & 42  &  480 & 5890  \\
   &(250,100)      &   21  &  160 & 28  &  210 & 2320 \\
500   &(150,140)      &   41  &  525 & 42  &  540 & 3870 \\
   &(150,150)      &   42  &  510 & 42  &  510 & 4275 \\
   &(200,150)      &   34  &  290 & 38  &  325 & 2870 \\
   &(250,150)      &   30  &  145 & 35  &  170 & 3815 \\
   &(200,200)      &   29  &  150 & 29  &  150 & 2050 \\
\hline
   &(300,150) & 41  & 195 & 44  & 210 &2023\\
   &(290,200) & 41  & 180 & 43  & 190 &1350\\
   &(300,250) & 38  & 130 & 42  & 140 &1056\\
800&(300,300) & 41  & 105 & 41  & 105 &1609\\
   &(350,350) & 40  &  40 & 40  &  40 &2632\\
   &(400,150) & 35  & 100 & 45  & 130 &2528\\
   &(400,200) & 37  &  85 & 45  & 105 &2430\\
   &(400,250) & 38  &  65 & 45  &  75 &1842\\
\hline
\end{tabular}
\end{center}
\caption{The $\HAbbtt$ and $\HAttbb$ signal efficiencies and the number of signal and total background events accepted into the final samples at $\sqrts$ = 500, 800~\GeV. The signal expectations 
are quoted for $\sin(\beta-\alpha)$ = 1 and the Higgs boson 
branching fractions of $\BrHbb$ = $\BrAbb$ = 90\%, 
$\BrHtt$ = $\BrAtt$ = 10\%.
\label{tab:sgnl_final_bbtt}
}
\end{table}

\begin{table}
\begin{center}
\begin{tabular}{|c|c|c|c|c|c|c|c|}
\hline
$\sqrts$ [GeV]&($\mH,\mA$) [GeV]   & \multicolumn{3}{c|}{$\mathrm{\Delta\mH}$ [GeV]} 
                        & \multicolumn{3}{c|}{$\mathrm{\Delta\mA}$ [GeV]}\\ \cline{3-8}
            &            & $\bbbb$ & $\bbtt$ & Comb.  
                        & $\bbbb$ & $\bbtt$ & Comb. \\
\hline \hline
   &(150,100)      & 0.12 & 0.21 & 0.10 & 0.12 & 0.18 & 0.10 \\
   &(200,100)      & 0.15 & 0.45 & 0.14 & 0.15 & 0.30 & 0.13 \\
   &(250,100)      & 0.27 & 0.55 & 0.24 & 0.27 & 0.46 & 0.23 \\
500&(150,140)      & 0.13 & 0.28 & 0.12 & 0.13 & 0.25 & 0.12 \\
   &(150,150)      & 0.12 & 0.40 & 0.11 & 0.12 & 0.40 & 0.11 \\
   &(200,150)      & 0.26 & 0.58 & 0.24 & 0.26 & 0.40 & 0.22 \\
   &(250,150)      & 0.48 & 1.40 & 0.45 & 0.48 & 0.77 & 0.41 \\
   &(200,200)      & 0.31 & 1.30 & 0.30 & 0.31 & 1.30 & 0.30 \\
\hline
   &(300,150)      & 0.33 & 0.44 & 0.26 & 0.33 & 0.57 & 0.29 \\
   &(290,200)      & 0.33 & 0.89 & 0.31 & 0.33 & 0.76 & 0.30 \\
   &(300,250)      & 0.45 & 1.08 & 0.42 & 0.45 & 1.02 & 0.41 \\
800&(300,300)      & 0.48 & 1.34 & 0.45 & 0.48 & 1.34 & 0.45 \\
   &(350,350)      & 0.97 & 2.12 & 0.88 & 0.97 & 1.12 & 0.73 \\
   &(400,150)      & 0.47 & 1.36 & 0.44 & 0.47 & 1.21 & 0.44 \\
   &(400,200)      & 0.74 & 2.00 & 0.69 & 0.74 & 1.88 & 0.69 \\
   &(400,250)      & 0.99 & 2.70 & 0.93 & 0.99 & 2.36 & 0.91 \\
\hline
\end{tabular}
\end{center}
\caption{
Precision on the Higgs boson mass determination for the $\bbbb$, $\bbtt$ channels and their combination at $\sqrts$ = 500 and 800~\GeV.
Numbers are quoted for $\sin^2(\beta-\alpha)$ = 1 and Higgs boson 
branching fractions of $\BrHbb$ = $\BrAbb$ = 90\%, 
$\BrHtt$ = $\BrAtt$ = 10\%.
\label{tab:prec_mass}
}
\end{table}

\begin{table}
\begin{center}
\begin{tabular}{|c|c|c|c|c|}
\hline
$\sqrts$ [GeV]&($\mH,\mA$) [GeV] & \multicolumn{3}{c|}{$\mathrm{\Delta\sigma/\sigma}$ [\%]} \\ \cline{3-5}
    &             & $\HAbbbb$ & $\HAbbtt$ & $\HAttbb$ \\
\hline \hline
   &(150,100)      & 1.6 &  5.2 &  4.9 \\
   &(200,100)      & 2.0 &  7.5 &  5.8 \\
   &(250,100)      & 2.4 & 16.3 &  8.9 \\
500&(150,140)      & 1.5 &  7.2 &  6.3 \\
   &(150,150)      & 1.5 &  4.2 &  4.2 \\
   &(200,150)      & 2.3 &  9.7 &  8.7 \\
   &(250,150)      & 3.5 & 22.6 & 13.5 \\
   &(200,200)      & 2.7 &  8.1 &  8.1 \\
\hline
   &(300,150)      & 2.7 & 10.0 &  8.8 \\
   &(290,200)      & 2.7 & 11.2 & 10.9 \\
   &(300,250)      & 3.0 & 13.8 & 11.9 \\
800&(300,300)      & 3.5 & 10.0 & 10.0 \\
   &(350,350)      & 6.6 & 17.0 & 17.0 \\
   &(400,150)      & 3.9 & 16.3 & 11.6 \\
   &(400,200)      & 4.3 & 21.5 & 19.8 \\
   &(400,250)      & 5.2 & 31.7 & 28.5 \\
\hline

\end{tabular}
\end{center}
\caption{
Relative uncertainty in the topological cross section measurements for the $\bbbb$, $\bbtt$ channels at $\sqrts$ = 500 and 800~\GeV.
Numbers are quoted for $\sin^2(\beta-\alpha)$ = 1 and the Higgs boson 
branching fractions of $\BrHbb$ = $\BrAbb$ = 90\%, 
$\BrHtt$ = $\BrAtt$ = 10\%.
\label{tab:prec_xsec}
}
\end{table}

\clearpage

\newpage

\begin{table}
\begin{center}
\begin{tabular}{|c|c|c|c|}
\hline
($\mH,\mA$) [GeV] & $\mathrm{\Gamma_{H,A}}$ [GeV] &  $\mathrm{\delta \gH}$ [GeV] & $\delta \gA$ [GeV] \\
\hline \hline
          & 0  & $<$0.9           & $<$0.9 \\ \cline{2-4}
(150,100) & 5  & $^{-0.6}_{+1.3}$ & $^{-0.6}_{+1.3}$\\ \cline{2-4}
          & 10 & $^{-1.3}_{+1.6}$ & $^{-1.1}_{+1.6}$ \\ \cline{2-4}
\hline \hline
          & 0  & $<$1.7           &   $<$1.8 \\ \cline{2-4}
(200,150) & 5  & $^{-1.1}_{+2.1}$ & $^{-1.2}_{+2.1}$ \\ \cline{2-4}
          & 10 & $^{-2.0}_{+2.6}$ & $^{-2.0}_{+3.0}$ \\ \cline{2-4}
\hline \hline
          & 0  & $<$3.6           & $<$3.6 \\ \cline{2-4}
(300,250) & 5  & $^{-2.5}_{+2.9}$ & $^{-2.5}_{+2.9}$ \\ \cline{2-4}
          & 10 &  $^{-2.9}_{+3.6}$&  $^{-2.8}_{+4.0}$ \\ \cline{2-4}
\hline
\end{tabular}
\end{center}
\caption{The accuracy of the Higgs boson width measurements 
in the case of large mass splitting between Higgs bosons.
The upper limits on the Higgs boson width in the case of 
$\gH$ = $\gA$ = 0 GeV correspond to 95\% confidence level.
\label{tab:width1}
}
\end{table}

\begin{table}
\begin{center}
\begin{tabular}{|c|c|c|}
\hline
($\mH,\mA$) [GeV] & $\mathrm{\Gamma_{H,A}}$ [GeV] & $\mathrm{\delta \Gamma_{H,A}}$ [GeV] \\
\hline \hline
          & 0  & $<$1.4  \\ \cline{2-3}
(150,150) & 5  & $^{-1.0}_{+1.4}$ \\ \cline{2-3}
          & 10 & $^{-1.7}_{+2.1}$ \\ \cline{2-3}
\hline \hline
          & 0  & $<$3.3 \\ \cline{2-3}
(200,200) & 5  & $^{-2.0}_{+2.5}$ \\ \cline{2-3}
          & 10 & $^{-3.2}_{+3.7}$ \\ \cline{2-3}
\hline \hline
          & 0  & $<$4.5 \\ \cline{2-3}
(300,300) & 5  & $^{-3.2}_{+3.7}$ \\ \cline{2-3}
          & 10 & $^{-4.0}_{+4.5}$  \\ \cline{2-3}
\hline
\end{tabular}
\end{center}
\caption{The accuracy on the Higgs boson width measurements in the case
of Higgs boson mass degeneracy.
The upper limits on the Higgs boson width in the case of 
$\gH$ = $\gA$ = 0 GeV correspond to 95\% confidence level.
\label{tab:width2}
}
\end{table}

\newpage


\vspace{-2cm}

\newpage
\clearpage

\newpage

\end{document}